\documentclass[12pt]{iopart}
\usepackage{amsbsy}
\usepackage{graphicx, epsfig, amssymb} 
\usepackage{bm} 

\usepackage[linktocpage]{hyperref}
\usepackage[caption=false]{subfig}
\usepackage[usenames]{color}
\usepackage{url}
\usepackage{color}

\newcommand{\f}{\frac}
\newcommand{\be}{\begin{equation}}
\newcommand{\ee}{\end{equation}}
\newcommand{\beq}{\begin{eqnarray}}
\newcommand{\eeq}{\end{eqnarray}}
\newcommand{\nn}{\nonumber}
\newcommand{\om}{\omega}

\begin{document}

\title{Slowly rotating thin shell gravastars}

\author{Nami Uchikata$^{1}$ and Shijun Yoshida$^{2}$}

\address{$^{1}$ Centro Multidisciplinar de Astrof\'\i sica --- CENTRA, Departamento de F\'\i sica, Instituto Superior T\'ecnico --- IST,
Universidade de Lisboa - UL, Av. Rovisco Pais 1, 1049-001 Lisboa, Portugal}
\address{$^{2}$Astronomical Institute, Tohoku University, Aramaki-Aoba, Aoba-ku, Sendai 980-8578, Japan}
 \eads{\mailto{nami.uchikata@ist.utl.pt}, \mailto{yoshida@astr.tohoku.ac.jp}}

\begin{abstract}
We construct the solutions of slowly rotating gravastars with a thin shell. In the zero-rotation limit, we consider the gravastar 
composed of a de Sitter core, a thin shell, and Schwarzschild exterior spacetime. The rotational effects are treated as small axisymmetric and stationary 
perturbations. The perturbed internal and external spacetimes are matched with a uniformly rotating thin shell. We assume that the angular velocity of 
the thin shell, $\Omega$, is much smaller than the Keplerian frequency of the nonrotating gravastar, $\Omega_k$. The solutions within an accuracy up to 
the second order of $\Omega/\Omega_k$ are obtained. The thin shell matter is assumed to be described by a perfect fluid and to satisfy the dominant 
energy condition in the zero-rotation limit. In this study, we assume that the equation of state for perturbations is the same 
as that of the unperturbed solution. The spherically symmetric component of the energy density perturbations, 
$\delta\sigma_0$, is assumed to vanish independently of the rotation rate.  Based on these assumptions, we obtain many numerical solutions and 
investigate properties of the rotational corrections to the structure of the thin shell gravastar. 
\end{abstract}

\submitto{\CQG}

\maketitle


\section{Introduction}
It is now widely believed that the black hole is a possible final state of massive stars after the gravitational collapse. 
Several tens of stellar-mass black hole candidates have been identified in our Galaxy. With a rough estimate, then, $10^8$ of stellar-mass black holes are expected to exist 
in our Galaxy. A supermassive black hole of $10^6$ to $10^9$ solar masses is also  believed to be in  the centers of most galaxies. 
Thus, it is likely that there is a huge number of black holes in the Universe. 
If quantum effects on gravity are taken into account, however, it could be that the gravitational collapse of objects comes to a halt 
and furthermore no event horizon forms~\cite{mar}. If this is the case, there is no black hole in the Universe. 
Some people have seriously considered this possibility and have proposed alternative models to the black hole that are almost 
as compact as black holes but have no event horizon. 

The first  serious model of such black hole-like objects is the so-called gravastars, gravitational vacuum stars, proposed by Mazur and 
Mottola~\cite{gravastar}. During the formation of the gravastar, just before the event horizon formation, a quantum vacuum phase transition 
is assumed to occur around the location where the event horizon would have been expected to form. As a result, the black hole event 
horizon formation is prevented. Actual equilibrium models of the gravastar are composed of a de Sitter core (a  positive cosmological 
constant matter core) surrounded by a relatively thin but finite-thickness shell whose equation of state is given by $p=\sigma$, where 
$\sigma $ and $p$ are 
the total energy density and pressure of the shell matter, respectively. By gluing these constituent parts, they consider objects having 
no event horizon nor spacetime singularity.  Mazur and Mottola also showed that their gravastars are thermodynamically stable 
unlike in the case of the black hole.

A simplified version of the gravastar model has been considered by Visser and Wiltshire~\cite{visser}. This simplified model is called 
the thin shell gravastar because the  relatively thin but finite-thickness shell of the original model is replaced by the single infinitesimally 
thin shell. They have further made detailed investigations of the dynamical properties of the thin shell gravastar and shown that 
there are thin shell gravastars stable against spherically symmetric perturbations. Some other models of the gravstar have been considered  
by many authors, see, e.g., \cite{carter,cattoen,cr,vitor1}.

In standard astronomical observations, the black hole candidates are basically chosen either because their estimated mass 
is greater than some theoretical limit, or on the basis of their strong external gravitational field. According to these astronomical criteria of 
the black hole candidate, therefore, all the black hole candidates could be alternative models whose mass and compactness are similar to those of 
the genuine black hole if the existence of such alternative models to the black hole is allowed theoretically. However, present and planned 
gravitational wave observatories will give us data with which one may test whether black hole candidates are genuine or not and check 
what is likely to be as an alternative model if the black hole candidates are not genuine black holes. 

One of such data is the characteristic oscillation frequency of the black hole-like object, which is called the quasinormal mode frequency 
because it is given by a complex number, whose imaginary part represents damping rate of the wave due to the energy loss. 
For the case of gravastars with a thick shell, Chirenti and Rezzolla~\cite {cr} have calculated quasinormal modes of axial parity perturbations. 
They showed that the decay rate of the gravastar quasinormal mode is substantially different from that of the black hole when the gravastar and 
back hole have the same gravitational mass. They also found no unstable eigenmode in the gravastars they considered. 
For the case of the thin shell gravastar, Pani {\it et al}.~\cite{pani} have 
calculated the quasinormal modes of the axial and polar parity perturbations. Their main results are the following: 
(i) The quasinormal frequency of the thin shell gravastar does not coincide with that of the  black hole even when 
the radius of the thin shell gravastar is nearly its Schwarzschild radius. This is because different boundary conditions at the nearly gravitational radius 
are required for the quasinormal modes of the gravastar and black hole. (ii) The thin shell gravastar may be distinguishable from 
the black hole by using the quasinormal frequencies. (iii) The particular models of the thin shell gravastars are unstable against nonradial  
perturbations for the case of less compact models. 

The other of such potential data obtaining from the gravitational wave observatories is the values of the lowest few multipole moments of 
the black hole-like object (its gravitational mass $M$, its spin angular momentum $J$, its mass quadrupole moment $Q$)~\cite{Ryan}. This data will be
obtained through the so-called extreme mass ratio inspiral (EMRI) associated with the supermassive black hole candidate in the center of a galaxy. 
Only when the central black hole-like object is non-spherically symmetric, one may obtain non-zero values of $J$ and $Q$. 
The quadrupole moment $Q$ includes information about the matter properties and is determined by $Q=J^2/M$ for the genuine black hole. Thus, one 
may, in principle, distinguish the genuine black hole from the alternative models to the back hole by using the values of $M$, $J$, and $Q$. 
 
So far, studies on the gravastar have mainly done for the spherically symmetric nonrotating case. As for the rotating gravastar, the first order rotational 
effect in the spin angular momentum $J$, which appears as the frame dragging effect, has been considered to investigate the so-called 
ergoregion instability of the gravastar~ \cite{vitor1,cr2}. Since most black hole candidates are expected to rotate rapidly because of 
conservation of angular momentum and/or matter accretion, it is necessary to extend the study of rotational effects on the gravastar structure to 
the higher order case. In this study,  we will therefore study rotating thin shell gravastars. As a first step, we focus on the case of slow rotation. 
We treat the rotational effects on the stellar structure as small perturbations about nonrotating spherically symmetric thin shell gravastar by 
using a method employed for investigation of the slowly rotating regular black hole~\cite{uchi}, which is based on the formalism to treat the slowly rotating 
star~\cite{hartle,thorne,chandra} and the matching of the distinct spacetimes~\cite{israel,ba}. The rotational effects are taken into account up to the second order 
in the spin angular moment $J$. Including the second order rotational effects, we may estimate the value of the mass quadrupole moment $Q$ of 
the gravastar, which is a quite important quantity to distinguish the black hole from the gravastar as mentioned before. 
For the nonrotating unperturbed solution, we use the thin shell gravastar model considered by Visser and Wiltshire~\cite{visser}.

The plan of this paper is the following: In section 2 we show the basic setup for matching of the two distinct spacetimes. 
The master equations for slowly rotating thin shell gravastars are given in section 3. 
In section 4, we show numerical results of the rotational corrections to the structure of the thin shell gravastar.
For the present gravastar models, the limit of the slowly rotating thin shell in vacuum asymptotically flat spacetime and the properties of 
the ergoreion are discussed in section 5. The conclusions are given in section 6. 
In this paper, we use the units of $G=c=1$, where $G$ and $c$ are the gravitation constant and the speed of light, respectively.
\section{Formulation: Matching of the two distinct spacetimes} 
\subsection{Spacetimes glued with a thin shell}
To obtain rotating models of the thin shell gravastar, in this study, we match two distinct spacetimes appropriate for them with a timelike thin shell. 
Following Visser and Wiltshire \cite{visser}, in the zero-rotation limit, the interior and exterior spacetimes of the thin shell are assumed to be 
the de Sitter and Schwarzschild spacetimes, respectively. We may obtain solutions of the nonrotating thin shell gravastar if the two spacetimes 
are properly matched at the coordinate radius $r=R$, where $2M<R<L$  with $2M$ and $L$ being the Schwarzschild and de Sitter horizon radii, 
respectively. Then, the thin shell gravastars, in general, have a massive spherical thin shell at $r=R$.

In this study, we shall restrict ourselves to the case of slow rotation, 
i.e., $\epsilon \equiv \Omega /\Omega_k \ll1$, where $\Omega$ is the angular velocity of the thin shell measured by an observer at infinity and $\Omega_k $ 
is the Kepler  frequency of the nonrotating thin shell gravastar. Following a standard technique developed for the analysis of the slowly rotating 
star~\cite{hartle,thorne,chandra},  we treat rotational effects on the stellar structure as stationary and axisymmetric perturbations around a zero-rotation 
solution. The rotational effects are taken into account up to an accuracy of $O(\epsilon^2)$ and we further assume that rotating thin shell 
gravastars have equatorial symmetry and time-azimuth reflection symmetry. Thus, the line elements for the interior and exterior spacetimes of the thin 
shell may  be written in the common form, given by 

\beq
ds ^2&=g_{\alpha\beta}\, dx^\alpha dx^\beta \nn \\
&= -f(r)(1+2 \epsilon^2 h(r ,\theta )) dt^2 + \f{1 }{f(r)}\left(1+\f{2 \epsilon^2 m(r,\theta )}{r f(r)}\right) dr^2 \nn \\
& + r^2(1+2\epsilon^2 k(r ,\theta )) \left[ d\theta ^2 +\sin^2 \theta \left\{d\phi  -\epsilon\,\om(r) dt \right\}^2 \right] O(\epsilon^3) ,
\label{in}
\eeq
where $g_{\alpha\beta}$ stands for the component of the metric tensor. Here and henceforth, the greek letter subscripts and superscripts denote 
indices running from $0$ to $3$ for four-dimensional tensor quantities. Outside and inside the thin shell, respectively, we assume that 
\beq
&(x^+)^{\mu} = (t^+,r^+,\theta^+, \phi^+), \nn \\
& f(r)=f^ +(r^+) = 1- {2M\over r^+} , \nn \\
&(h(r,\theta),k(r,\theta),m(r,\theta),\om(r)) =  \nn \\
& \quad (h^+(r^+,\theta^+),k^+(r^+,\theta^+),m^+(r^+,\theta^+),\om^+(r^+)), 
\eeq
and  
\beq
&(x^-)^{\mu} = (t^-,r^-,\theta^-, \phi^-), \nn \\
& f(r)=f^ -(r^-) = 1- {(r^-)^2\over L^2} , \nn \\
&(h(r,\theta),k(r,\theta),m(r,\theta),\om(r)) =\nn \\
& \quad (h^-(r^-,\theta^-),k^-(r^-,\theta^-),m^-(r^-,\theta^-),\om^-(r^-)).
\eeq
Here and henceforth, the quantities with the superscripts $(^+)$ and $(^-)$ mean the quantities defined outside and inside the thin shell, respectively. 
For the sake of simplicity, however, the superscripts $(^+)$ and $(^-)$ are sometimes omitted when it does not cause any confusion. 
We further assume that 
\beq
& h(r,\theta)=h_0(r)+h_2(r)P_2(\cos\theta), \nn \\
& m(r,\theta)=m_0(r)+m_2(r)P_2(\cos\theta), \nn \\
& k(r,\theta)=k_2(r)P_2(\cos\theta).
\eeq

For the exterior spacetime of the thin shell, the functions $\om^+$, $h^+$, $k^+$, and $m^+$ are the solutions regular at infinity obtained by 
solving the vacuum Einstein equations, see, e.g., \cite{uchi,hartle}, given by 
\beq
&\om ^+  = \f {2 J} {r^3} ,\\
& m_0^+ =\delta M -\f{J^2 }{r^3}, \\
& h_0 ^+= - \f{\delta M}{r- 2M } + \f{J^2 }{r^3 (r- 2M)}  ,\\
& h _2 ^+= J^2 \left (\f{1}{ M r^3} + \f{1}{r^4}\right ) +  B Q_2 ^{\, 2} \left (\f{r}{M} -1 \right ) ,\\ 
& k _2 ^+=- \f{J^2}{  r^4} -  B\f{2 M}{ \sqrt {r (r-2 M)}} Q_2 ^{\, 1} \left (\f{r}{M} -1 \right )-h_2^+ ,\\
& m_2  ^+= (r-2 M)  \left (- h_2 ^+ +\f{r^4}{6}\left (\f{d\om}{dr} \right ) ^2 \right ), 
\eeq
where $\delta M$ means change in the total gravitational mass due to the rotation, $J$ the total angular momentum, $Q_l^{\, m}$ 
the associated Legendre function of the second kind, and $B$ an integral constant. The integral constant $B$ is related to the quadrupole moment 
of the gravastar $Q$ through the relation  $Q =J^2/ M +8 B M^3 /5$ (see, e.g., \cite{hartle}). Note that the case of $B=0$ corresponds to the slowly 
rotating Kerr solution. The associated Legendre functions $Q_2^{\, 2}$ and $Q_2^{\, 1}$ are, respectively, defined by 
\beq
Q_2^{\, 2}(x) & = \f{x(5-3 x^2)} {x^2 -1} +\f{3(x^2 -1)}{2}  \log \f {x+1} {x-1},\\
Q_2^{\, 1}(x)& = \sqrt{x^2 -1 } \left (\f{2-3x^2}{x^2-1}+ \f {3 x}{2} \log \f {x+1} {x-1} \right ) .
\eeq

For  the interior spacetime of the thin shell, the functions $\om^-$, $h^-$, $k^-$, and $m^-$ are the solutions regular on the symmetry axis 
obtained by solving the Einstein equation with a positive cosmological constant, see, e.g., \cite{uchi}, given by 
\beq
& \om ^ -= C_1,\\ 
& m_0 ^ -=0, \\
& h_0 ^-=C_2 ,\\
& h _2^- = \f { C_3}{8 r^2} \left (\f {-3 L^2 + 5  r^2} { L^2  f^ - (r) } + \f {3 L f^-(r) \mbox{Arctanh} (r/L)}{r} \right ), \label{h2n}\\ 
& k _2 ^-=\f { C_3}{8 r^2 L} \left (\f {3 L^2 + 4  r^2} { L } - \f {3 (L^2+r^2) \mbox{Arctanh} (r/L)}{r} \right ), \label{k2n} \\
& m_2  ^-= -r f^ -(r) h_2 ^-, 
\eeq
where $C_1$, $C_2$, and $C_3$ are integral constants.

\subsection{First junction conditions}
In order to match the two distinct spacetimes given in the previous subsection properly, the so-called Israel's junction conditions~\cite{israel,ba} 
have to be fulfilled.  Israel's junction conditions are composed of two parts. The first conditions are related to the induced metrics on the thin 
shell, ${h^\pm}_{ab}$, and the second conditions to the extrinsic curvatures on the thin shell, ${K^\pm}_{ab}$. Here, let us consider the first 
conditions, which require that ${h^\pm}_{ab}$ is continuous across the thin shell, i.e., 
\be
 [[h_{ab}]] =0,
\label{junc1}
\ee
\be
h_{ab}=g _{\alpha \beta} e^{\alpha} _a e ^{\beta} _b \,, \quad 
e^{\alpha} _a =\f {\partial x^{\alpha}} {\partial y^a},
\ee
where $e^{\alpha} _a$ means the tangent vectors to the thin shell, $\Sigma$, and $y^a$ denotes the intrinsic coordinate of the thin shell, whose 
equation is given in the form $x^{\alpha}=x^{\alpha}(y^a)$. 
Here and henceforth, the double square brackets $[[ q]]$ imply the jump in the quantity $q$ on the thin shell, i.e., $[[q]] \equiv q | ^+_\Sigma -q | ^-_\Sigma$, 
and the roman letter subscripts and superscripts denote indices for the three-dimensional tensor quantities defined on the thin shell.
In this study, the intrinsic coordinate of the shell is written by $y^a = (T, \Theta, \Phi)$ and the equations of the thin shell in the interior and exterior 
spacetimes of the thin shell are assumed to be  
\be
(x^{\pm})^{\mu} = (A^{\pm} T, R + \epsilon ^2\xi^{\pm}(\Theta), \Theta+\epsilon^2 (l^{\pm})^{\Theta}(\Theta), \Phi)+O(\epsilon^3), 
\ee
where $R\equiv R^\pm$ is the coordinate radius of the thin shell in the zero-rotation limit.  In this study, we may further assume that $A^+ = 1$, 
$A^-={\rm constant}$, and $(l^+)^{\Theta} = 0$ by using the freedom in the coordinate choice (see \cite{uchi}). The tangent vectors to $\Sigma$ 
may  be then given by 
\beq
(e^{\mu}_{T})^{\pm}&= \left (A^{\pm} , 0 ,0, 0 \right)+ O(\epsilon^3), \nn \\
(e^{\mu}_{\Theta})^{\pm}&= \left (0 , \epsilon^2 \xi^{\pm}_{,\Theta} ,1+\epsilon^2(l^{\pm})^{\Theta}_{, \Theta}, 0 \right)+ O(\epsilon^3), \nn \\
(e^{\mu}_{\Phi} )^{\pm}&= \left (0 , 0 ,0, 1 \right) + O(\epsilon^3).
\eeq
To achieve separation of variables, we assume that  
\be
\xi (\Theta) = \xi_0 + \xi_2 P_2(\cos \Theta), 
\ee
where $\xi_0$ and $\xi_2$ are constants. 

The $\epsilon^0$- and $\epsilon$-order first junction conditions, $\displaystyle \lim_{\epsilon \rightarrow 0} [[h_{TT}]]=0$ and 
$\displaystyle \lim_{\epsilon \rightarrow 0} \partial_\epsilon [[h_{T\Phi}]]=0$, respectively, lead to
\beq
&(A^-)^2 = \f {f^+(R)} {f^-(R)} , \quad  \om ^- =C_1 = \f {2J} {R^3A^-} . 
\label{match1}
\eeq
The conditions $[[h_{\Theta \Theta} ]]=0$ and $[[h_{\Phi \Phi } ]]=0$ lead to
\beq
[[\xi _ 0 ]]- R (l^-)^{\Theta}_{\, , \Theta} = 0, \, \, [[\xi _ 0 ]]= 0,\,\, \left [\left [\f { \xi_2} {R} + k_2  (R)\right ]\right ]= 0.
\label{match2}
\eeq
We therefore have $(l^-)^{\Theta}=0$ because of the first two equations in (\ref{match2}) and the regularity on the symmetry axis. 
From the $\epsilon^2$-order conditions of $\displaystyle \lim_{\epsilon \rightarrow 0} {1\over 2}\partial^2_\epsilon[[h_{TT}]] = 0$, we obtain 
\beq
&  [[h_0 (R)]] +\f {R \xi^-_0} {L^2 f^-(R)} + \f {M \xi^+_0} {R^2 f^+ (R)} =0,  \nn \\
& [[h_2 (R)]] +\f {R \xi^-_2} {L^2 f^-(R)} + \f {M \xi^+_2} {R^2 f^+ (R)} =0. 
\label{match3}
\eeq
The conditions  $[[h_{T\Theta}]]=0$ and $[[h_{\Theta \Phi}]]=0$ are automatically satisfied because they are smaller than $\epsilon^3$-order 
quantities. Nonzero components of the induced metric may be, in terms of quantities associated with the exterior spacetime of the thin shell, 
given by 
\begin{eqnarray}
h_{TT}&=&-f^+ -\epsilon^2 f^+ \left[  -R^2  \sin ^2\Theta (\omega^+)^2  +(\ln f^+)' \xi^+_0+2h^+_0  \nonumber  \right. \\
&&\left. +\left\{ (\ln f^+)' \xi^+_2+2h^+_2  \right\} P_2 \right] + O(\epsilon^4)  \,,  \label{hTT} \\
h_{\Theta\Theta}&=&R^2+\epsilon^2  2 R \left[ \xi^+_0+(R k^+_2 +\xi^+_2)P_2 \right] + O(\epsilon^4) \,, \\
h_{\Phi\Phi}&=&\sin ^2\Theta h_{\Theta\Theta} \,, \\
h_{T\Phi}&=&h_{\Phi T}=-\epsilon R^2 \omega^+  \sin ^2\Theta + O(\epsilon^3) \,, 
\end{eqnarray}
where $q^\prime$ means $d q/dr |_{r=R}$. 

In summary, the first junction conditions give us the six constraint conditions, the six conditions given in equations (\ref{match1})--(\ref{match3}), 
for the eleven constants $J$, $\delta M$, $B$, $C_1$, $C_2$, $C_3$, $A^-$, $\xi^\pm_0$, and $\xi^\pm_2$. Thus, the first junction 
condition, (\ref{junc1}), can be fully satisfied as shown later.

\subsection{Second junction conditions}
While the first junction conditions, (\ref{junc1}), ensure that the two distinct spacetimes are joined smoothly, 
the second junction conditions give us information about the matter distribution on the matching surface $\Sigma$ in 
terms of the jump in the extrinsic curvature across $\Sigma$ (see, e.g., \cite{israel,ba}). According to Israel's junction conditions,  the stress-energy 
tensor of the thin shell $S_{ab}$ is given by 
\be
S_{ab} = {1\over 8\pi}\left([[ K_{ab} ]] -h_{ab} [[K]] \right),
\label{sab}
\ee
\be
K_{ab} \equiv  - n _{\alpha ; \beta} e^{\alpha} _a e ^{\beta} _b \quad \mbox{and} \quad  K=h_{ab} K^{ab}\,,
\ee
where $n^\alpha$ is the unit normal vector to $\Sigma$, for which $n_\alpha e^\alpha_a=0$ and $n^{\mu}n_{\mu} = 1$, and the semicolon 
$(;)$ denotes the covariant derivative associated with $g_{\alpha\beta}$. Nonzero components of $n_{\mu}$ are given by 
\beq
n_r & =\f{1}{\sqrt{f}}+\epsilon^2 \f{m_0+m_2 P_2(\cos \theta)}{r\sqrt{f}^3} +O(\epsilon ^4),\\
n_{\theta} & = -\epsilon ^2 \f{\xi_2}{\sqrt{f}} \partial_\theta P_2 (\cos \theta)+O(\epsilon ^4).
\eeq
Nonzero components of the extrinsic curvature up to an accuracy of $O(\epsilon)$ are then given by 
\beq
& K^T_T= -\f{f^{\prime}}{2\sqrt{f}} + O(\epsilon^2) , \\
& K^{\Theta}_{\Theta}=K^{\Phi}_{\Phi}= - \f{\sqrt{f}} {R} + O(\epsilon^2), \\
& K^{\Phi}_T =\epsilon \f{ f (2 \om + R \om ^{\prime})-R \om f^{\prime} }{2 R \sqrt{f} } A+ O(\epsilon^3) ,\\
& K^T_{\Phi} =-\epsilon \f{R^2 \om^{\prime}}{2 A \sqrt{f} } \sin^2 \Theta + O(\epsilon^3) , \\
& K =- \f{4f+Rf^{\prime}}{2R\sqrt{f}} + O(\epsilon^2). 
\eeq
Explicit forms of the $\epsilon^2$-order coefficients of the extrinsic curvature component are summarized in Appendix A.

The matter three-velocity tangent to $\Sigma$, $u^a$, and the total energy density measured by an observer with $u^a$, $\sigma$, are, 
respectively, defined  by the timelike eigenvector and the corresponding eigenvalue of $S^a_b$, given by 
\be
S^a _{\>\, b}u ^b  = - \sigma u^a, \quad u^a u_a = -1, 
\label{su}
\ee
where $u^a$ is related to the matter four-velocity associated with the thin shell through $u^\alpha=e^\alpha_a u^a$.
The stress tensor of the thin shell, $\gamma _{ab}$, measured by an observer with $u^a$, is defined by
\be
\gamma _{ab} = S^{cd} q_{ac} q_{bd},
\ee
where $q_{ab}$ is the projection tensor orthogonal to the $u^a$, defined by $q_{ab} =h_{ab}+u_a u_b $.
For a perfect fluid, $\gamma_{ab}$ is proportional to $q_{ab}$, $\gamma _{ab} = p\,q_{ab}$, and $p$ is interpreted as the isotropic pressure of the fluid.
\section{Master equations for slowly rotating thin shell gravastar} 

\subsection{Non-rotating thin shell gravastars}
In the limit of $\epsilon\rightarrow 0$, (\ref{su}) may be solved if we assume that $\displaystyle u^a={1\over \sqrt{f^+}}\,\delta^a_T+O(\epsilon)$ with 
$\delta^a_b$ being the Kronecker delta. Thus, we obtain the energy density and pressure in the zero-rotation limit as follows: 
\beq
&\sigma_0\equiv\lim_{\epsilon\rightarrow 0}\sigma=\f{\sqrt{f^ -} - \sqrt{f ^+} }{4 \pi R}, \label{sig00} \\
& p_0 \equiv\lim_{\epsilon\rightarrow 0}p= - \f{1}{8 \pi R^2} \left ( \f{M-R}{\sqrt{f ^+}} + R\f{1-{2R^2\over L^2} }{\sqrt{f ^-}} \right ). 
\label{p00}
\eeq
Note that the relation $\gamma _{ab} = p\,q_{ab}+O(\epsilon^2)$ may be trivially confirmed in the present situation. In order to specify an equilibrium 
solution of the nonrotating thin shell gravastar, following \cite{visser}, we give values of two dimesionless parameters $R/M$ and $M/L$ under 
the condition of $2M<R<L$.  Then, we may obtain nondimensional fundamental physical quantities for the nonrotating thin shell gravastar,  $A^-$, 
$4\pi\sigma_0 R^2/M$, and $p_0/\sigma_0$, by using the first equation in (\ref{match1}), (\ref{sig00}) and (\ref{p00}). 

\subsection{$\epsilon$-order rotational effects: frame dragging}

In this study, we focus on the case of the uniformly rotating thin shell around the symmetry axis. Thus, $u^a$ is given by  
\be
{u^{\Phi} \over  u^T}={d\Phi\over dT}=\Omega={\rm constant}\,, \quad u^\Theta=0\,, \label{SeigenP}
\ee
where $\Omega$ is the angular velocity of the thin shell and $\Omega=O(\epsilon)$. As a result, the $\Theta$ component of (\ref{su})  is 
automatically satisfied since $S^{\Theta}_{\Phi} =O(\epsilon^3)=S ^{\Theta}_{T}$, and the $T$ and $\Phi$ components of (\ref{su}) become
\be
S^{T} _T + S^{T} _{\Phi} \Omega= -\sigma +O(\epsilon^4)\,,
\label{Stt}
\ee
and
\be
S^{\Phi} _T+ S^{\Phi} _{\Phi} \Omega= -\sigma_0 \, \Omega+O(\epsilon^3)\,, 
\label{Stp}
\ee
respectively. Equation~(\ref{Stp}) leads to the relation between the angular velocity of the thin shell $\Omega$ and the total angular moment $J$, 
given by 
\beq
\Omega &   \equiv \epsilon \Omega_k \nn \\
&= -  \f{S ^{\Phi} _{\>\, T}}{S^{\Phi}_{\> \Phi} +\sigma_0} +O(\epsilon^3) \nn\\
&= \epsilon \f{J}{R^2}   \f{\sqrt{f ^-}+ 2   \sqrt{f ^+} }{ \sqrt{f^+} R  - (R -3M)   \sqrt{f^-}} \,, 
\label{defOmega}
\eeq
or 
\be
J = R^2\f {\sqrt{f^+} R  - (R -3M)   \sqrt{f^-}} { \sqrt{f ^-}+ 2   \sqrt{f ^+} } \Omega_k.
\ee
Here, as mentioned before, $\Omega_k$ is the Kepler frequency of the nonrotating thin shell gravastar, defined by $\sqrt{M/R^3}$. 
Equation~(\ref{Stt}) determines the total energy density up to an accuracy of $O(\epsilon^2)$. As seen from the above, we confirm that $u^a$ 
given in (\ref{SeigenP}) is a timelike eigenvector of $S^a_b$ within an accuracy up to $O(\epsilon^3)$. The angular velocity of the frame 
dragging in the interior spacetime of the thin shell $\omega^-$ is, in terms of $J$, given by the second equation in (\ref{match1}).    

\subsection{$\epsilon^2$-order rotational effects: deformation of the matter distribution due to rotation} 

As shown in the previous subsection, the three-velocity of the thin shell $u^a$ is given by (\ref{SeigenP}). Then, the corresponding 
eigenvalue $\sigma$ is given by (\ref{Stt}). $\sigma$ is the total energy density and may be written in the form, given by 
\beq
\sigma&=&\sigma_0+\epsilon^2 \delta \sigma +O(\epsilon^4) \nn \\
&\equiv&\sigma_0+ \epsilon^2(\delta \sigma_0 + \delta \sigma_2 P_2)+O(\epsilon^4)  \,, 
\eeq
where the explicit forms of $\delta\sigma_0$ and $\delta\sigma_2$ are given in Appendix B. 
While $\Omega$ is related to $J$ through (\ref{defOmega}), as mentioned before, $u^T$ is determined by the normalized condition 
of $u^a$, 
\be
h_{ab} u^a u^b = -1 \,.
\ee 
Thus, we have 
\beq
u^T = {1\over\sqrt{f^+}} &+ \epsilon ^2 \left(\f {-(f^+)^{\prime} \xi^+(\Theta)+ R^2(\Omega_k - \om^+)\sin ^2\Theta} {2 \sqrt{f^+}^3} -\f {h^+(R, \Theta)} {\sqrt{f^+}}\right ) \nn \\
&+ O(\epsilon ^4)\,,
\eeq
\be
u^\Phi={1\over\sqrt{f^+}}\Omega  + O(\epsilon ^3) \,. 
\ee
By using these expressions for $u^a$, we may obtain nonzero components of the stress tensor of the thin shell, $\gamma^a_{\>b}$, given by
\beq
\gamma^T _{\> T} &= p_0 \, q^T _{\> T}+O(\epsilon^4), \quad \gamma^T _{\> \Phi} =p_0\,  q^T _{\> \Phi} O(\epsilon ^3), \quad \gamma^{\Phi} _{\> T} &=p_0 \, q^{\Phi} _{\> T}+O(\epsilon^3) ,
\label{trigam}
\eeq
\beq
\gamma^+ & = p_0  -\f {\epsilon ^2 } {8 \pi R^2} \left \{   \f {2(\Omega_k R^3-2 J)} {3  R^3 f^+} \left (\f{JL^2 -R^5 \Omega_1}{L^2 \sqrt{f^-}} \right.  +\f{ J(3 M -2 R)+M R^3 \Omega_k}{R \sqrt{f^+}} \right ) \nn \\
&-R^2 \sqrt {f^+} (h_0 ^+)^{\prime} -\f {\xi^-_0} {\sqrt{f^-} ^3}  \left.+ \f {(3 M^2 -3 M R + R^2) \xi^+_0 +R(R-M) m_0 ^+} {R^2 \sqrt {f^+}^3} \right \}\nn \\
&+ \f {\epsilon ^2 P_2} {8 \pi R^2} \left \{ \f {2(J L^2 -R^5 \Omega_k)(R^3 \Omega_k -2 J^2) } {3L^2 R^3\sqrt{f^-} f^+}\right.    + R^2[[ \sqrt {f} ( h_2  ^{\prime} + k_2 ^{\prime})]] \nn \\
&+\f {(3 R^2 - 2 L^2) \xi^-_2 +(L^2 - 2 R^2 ) m_2 ^-} {L^2 \sqrt {f^-} ^3}  +\f{2(2 J -R^3 \Omega_k) (J(3M-2 R) +M R^3 \Omega_k) }{3 R^4 \sqrt{f^+}}  \nn \\
& \left . - \f {(3 M^2 + 3 M R -2 R^2) \xi^+_2 + R(R-M) m_2 ^+} {R^2\sqrt {f^+}^3} \right \}  + O(\epsilon ^4 ),
\eeq
\beq
 \gamma^- & = \f {\epsilon ^2 \sin ^2 \theta} {16 \pi R^2} \left \{ \f {2 (2 J - R^3 \Omega_k)} { R^3 f^+} \left (   -\f {JL^2 -R^5 \Omega_k} {L^2 \sqrt{f^-} } \right. \right. \nn \\
 & \left .\left. +\f {J(3 M -2 R) + M R^3 \Omega_k} {R\sqrt{f^+}}\right ) +3 \left[ \left[ \f {\xi_2 } {\sqrt {f}} \right ] \right] \right \}   + O(\epsilon ^4 ),
 \label{gam_mai}
\eeq
where $\gamma^{\pm} = (\gamma^{\Theta} _{\>\Theta} \pm \gamma^{\Phi} _{\>\Phi})/2$.

In this study, we focus on the thin shell composed of a perfect fluid. For a perfect fluid, as mentioned before, the stress tensor satisfies the relation 
$\gamma_{ab} = p \, q_{ab}$. In equations (\ref{trigam})-(\ref{gam_mai}), we see that the relation $\gamma_{ab} = p \, q_{ab}$ is satisfied within an 
accuracy up to $O(\epsilon^3)$ except for $\gamma ^{\Theta} _{\> \Theta} $ and $\gamma^{\Phi} _{\>\Phi}$. If the thin shell matter is described as 
a perfect fluid, $\gamma^{\Phi} _{\>\Phi}$ and $\gamma ^{\Theta} _{\> \Theta} $ are written as 
\beq
 \gamma^{\Phi} _{\>\Phi} &= p\, q^{\Phi} _{\>\Phi} \nn \\
 &= (p_0 +\epsilon^2 \delta p) (1 +u^{\Phi}u_{\Phi})+O(\epsilon^4) \\
 &= p_0 + \epsilon^2 \delta p +p_0 u^{\Phi}u_{\Phi}+O(\epsilon^4) \nn, 
\eeq
and 
\beq
 \gamma ^{\Theta} _{\> \Theta}  &= p\,q^{\Theta} _{\> \Theta}  = p_0 +\epsilon^2 \delta p +O(\epsilon^4) .
\eeq
Thus, the condition for the thin shell to be composed of a perfect fluid is given by 
\be
\gamma^{\Phi} _{\>\Phi}-\gamma^{\Theta} _{\>\Theta} = p_0 u^{\Phi}u_{\Phi}.
\label{pful}
\ee
This condition may be rewritten in the form, given by 
\beq
\left [ \left[\f{\xi_2}{ \sqrt{f}} \right] \right ]& =\f{2}{3 \sqrt{f^+}} \left (\f{4 J^2}{R^3} - \f{2 J  (M+R) \Omega_k}{R} +  M R^2 \Omega_k ^2 \right ) \nn \\
&-2(2 J -R^3 \Omega_k) \left ( \f {4\pi R p_0} {3}  \Omega_k -\f{ J L^2 - R^5 \Omega_k }{3 L^2 R^3\sqrt{f^-}}   \right ) .
\label{xi2}
\eeq
The pressure of the thin shell $p$ may be given by
\beq
p&=&\gamma ^+ -\f{1}{2} \, p_0 u^{\Phi}u_{\Phi} \\
&\equiv& p_0+\epsilon^2\delta p  \\
&\equiv& p_0+\epsilon^2(\delta p_0 + \delta p_2 P_2) \,, 
\eeq
where the explicit expressions for $\delta p_0$ and $\delta p_2$ are summarized in Appendix B. 

Since the perfect-fluid thin shell is considered, the total energy and pressure perturbations have to be related by the relationship, 
\be 
\delta p={dp\over d\sigma}\, \delta\sigma \,, 
\label{dpdsig}
\ee
where $\displaystyle {dp\over d\sigma}$ is a function of the thermodynamic quantities. This implies that 
\be
\delta p_0={dp\over d\sigma}\, \delta\sigma_0 \,, \quad  \delta p_2={dp\over d\sigma}\, \delta\sigma_2 \,. 
\label{eosP}
\ee
Once the coefficient $\displaystyle {dp\over d\sigma}$ is known, we have two more equations for determining the eight constants $\delta M$, $B$, 
$C_2$, $C_3$,  $\xi^\pm_0$, and $\xi^\pm_2$ associated with $\epsilon^2$-order rotational effects.

The second equation in (\ref{eosP}), (\ref{xi2}), the second equation in (\ref{match3}), and 
the third equation in (\ref{match2}) are composed of  four coupled linear algebraic equations among the four constants $B$, $C_3$, 
and $\xi^\pm_2$ associated with $\epsilon^2$-order quadrupole perturbations. Thus, we may obtain solutions of the quadrupole perturbations 
by solving these coupled linear algebraic equations (see Appendix C). 

To determine the four constants $\delta M$, $\xi_0^\pm$, and $C_2$ associated with the spherically symmetric perturbation, 
an additional condition is required because we have so far obtained the three algebraic equations;  
the first equation in (\ref{eosP}), the first equation in (\ref{match3}), and the second equation in (\ref{match2}).  
As a fourth condition, in this study, we impose that $\delta \sigma_0 = 0$. By adding this condition, we may uniquely determine 
the four constants $\delta M$, $\xi_0^\pm$, and $C_2$. 

From the first law of thermodynamics, changes in the particle number density of the thin shell $\delta n$ is given by 
\be
\delta n= \f{n_0}{\sigma_0 +p_0}\delta \sigma,
\ee
where $n_0$ is the number density in the zero-rotation limit. Changes in the total particle number of the thin shell $\delta N$ is given by
\beq
\delta N &=& n_0 \int \left (\f{\delta \sigma}{\sigma_0 +p_0}u^T+\delta u^T   +u^T{\delta \sqrt{-h}\over \sqrt{-h}} \right)\sqrt{-h} \, d \Theta d \Phi,
\eeq
where $h$ means the determinant of $h_{ab}$. This leads to
\be
\f{ \delta N } {N} = \f{2 \xi^+_0} {R} + \f {(2 J - R^3 \Omega_k)^2  } {3 R^4 f^+} \,,
\label{dN0}
\ee
where $N$ denotes the total particle number of the nonrotating thin shell gravastar. 
Note that $\delta N$ is given only by the spherically symmetric perturbation quantities, because the terms related to the quadrupole perturbations 
vanish after the angular integration.

\section{Numerical results}

\subsection{Thin shell gravastar in the zero-rotation limit}

\begin{figure}[t]
\begin{center}
\includegraphics {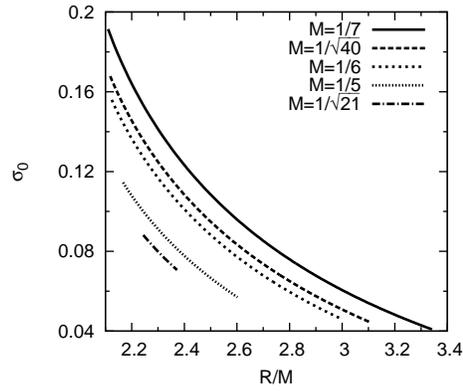}
\caption{The energy density of the thin shell $\sigma_0$ in the zero-rotation limit as a function of $R/M$.}
\label{sig0}
\end{center}
\end{figure}
\begin{figure}
\begin{center}
\includegraphics {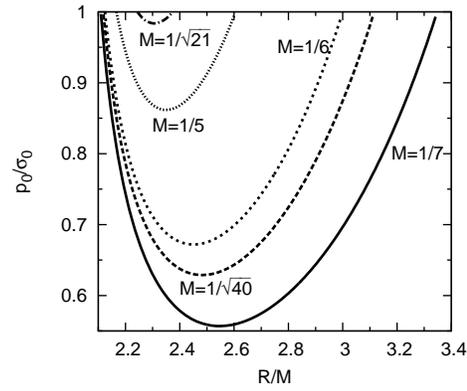}
\caption{The same as figure \ref{sig0} but for the ratio of the pressure of the shell to the energy density of the thin shell $p_0/\sigma_0$.}
\label{p0s0}
\end{center}
\end{figure}
\begin{figure}
\begin{center}
\includegraphics {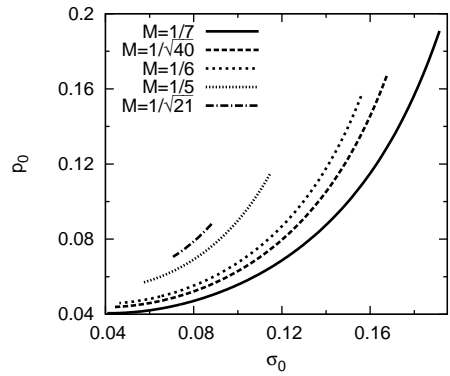}
\caption{The pressure of the thin shell $p_0$ in the zero rotation limit as a function of the energy density of the thin shell $\sigma_0$.}
\label{eos0}
\end{center}
\end{figure}
\begin{figure}
\begin{center}
\includegraphics {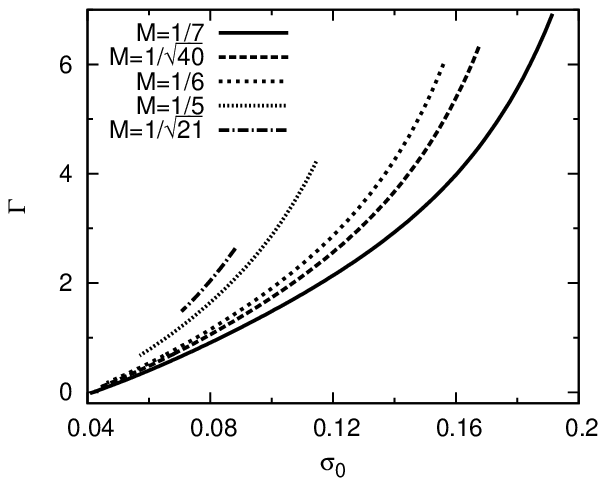}
\caption{The adiabatic index $\Gamma$ for the thin shell matter as a function of the energy density of the shell $\sigma_0$.}
\label{dps0}
\end{center}
\end{figure}
Mazur and Mottola have proposed the original gravastar model, in which a finite-thickness shell described  by the stiff equation of state, 
$p=\sigma$, is assumed to be around the radius at which the horizon would have formed in the standard classical general relativity~\cite{gravastar}. 
A simplified version of the gravastar model has been considered by Visser and Wiltshire~\cite{visser}.
This simplified model is called the thin shell gravastar because the finite-thickness shell of the original model is replaced by the single thin shell. 
They have made detailed investigations of the properties of the thin shell gravastar and shown that there are gravastar models stable against 
spherically symmetric perturbations. If spherically symmetric nonrotating models are unstable, to study rotational effects on such models will 
lose its physical significance. In this study, therefore, we only consider the stable thin shell gravastar models as solutions in the zero-rotation limit.
The stable nonrotating thin shell gravastar models that we employ in this study correspond to the case of $V=0$ in \cite{visser}. 

As mentioned before, we assume that $2M < R < L$ to avoid appearing solutions with horizons. Following Visser and Wiltshire \cite{visser}, 
we require that the thin shell matter satisfies the weak and dominant energy conditions. The weak energy condition $\sigma_0 \ge 0$ and \ref{sig00} lead to the constraint equation, given by
\be
\f{R}{M} \le \left (\f{L^2}{2 M^2} \right )^{\f{1}{3}}. 
\label{ineq0}
\ee
Thus, we have 
\be
2 < \f{R}{M} \le \left (\f{L^2}{2 M^2} \right )^{\f{1}{3}}\,. 
\label{ineq1}
\ee
Although it is a complicated numerical task to translate the dominant energy condition for the thin shell matter, $\sigma_0 \ge |p_0|$, to the condition given 
explicitly in terms of $R$, $M$, and $L$, a simultaneous analysis of the dominant energy condition and the radial stability condition can derive the condition, 
\be
{M^2\over 2L^2} \lesssim 0.024304 \, , 
\label{ineq2}
\ee
for the thin shell to satisfy the both conditions (see, for detailed derivation, \cite{visser}). In this study, therefore, we consider solutions of the thin 
shell gravastar in the zero-rotation limit characterized by (\ref{ineq1}) and (\ref{ineq2}). 

To investigate rotational effects on the structure of 
the thin shell gravastar, changing values of $R/M$, we obtain the five sequences of equilibrium solutions of the spherical gravastar characterized by the fixed 
values of $M/L$, which are chosen as the following: $M/L = 1/\sqrt{21}$, $1/5$, $1/6$, $1/\sqrt{40}$, and $1/7$. For showing numerical results of the 
gravastar, in this paper, we will use the units of $L=1$. In figures \ref{sig0} and \ref{p0s0}, the total energy density $\sigma_0$ and the ratio of the pressure 
to the total energy density $p_0/\sigma_0$ of 
the thin shell for the equilibrium sequences of the gravastar are given as functions of $R/M$, respectively. In these figures, each line corresponds to 
an equilibrium sequence characterized by the fixed values of $M/L$ whose values are given in these figures. In these figures, we see that the energy density of 
the shell $\sigma_0$ is a decreasing function of $R/M$. 
As can be seen from figure \ref{p0s0}, the endpoints of each sequence considered in this study correspond to the solutions whose thin shells are described by the stiff equation of state, $p_0=\sigma_0$. For a sequence of  equilibrium solutions characterized by a fixed value of $M/L$, therefore, there 
are two solutions with the stiff thin shell. Although in all the solutions shown in figures \ref{sig0} and \ref{p0s0}, the thin shells satisfy the weak and dominant energy 
conditions and are stable against radial perturbations, i.e., they are equally plausible, thin shell gravastar solutions having smaller radius would be more relevant 
because they should be black hole-like objects.  Note that most unperturbed solutions we consider have their radius of $R/M <3$. Such solutions might be 
nonlinearly unstable as pointed out in \cite{jk,vitor3}. 

Considering the equilibrium sequences characterized by fixed values of $M/L$, we may obtain equations of state for the thin shell matter, 
$p_0=p_0(\sigma_0)$, depending on values of $M/L$. These equations of state are shown in figure \ref{eos0}. By assuming that these equations 
of state specify values of $\displaystyle{dp\over d\sigma}$ appearing in (\ref{dpdsig}), we may reasonably determine the equation of state 
for the perturbation. We therefore assume that 
\be
\f{dp} {d \sigma} \equiv \f {{\partial p_0 \over \partial R}} {{\partial \sigma_0 \over \partial R}} = -\f {R} {2 (\sigma_0 +p_0)} \f {\partial p_0} {\partial R} .
\label{eos3}
\ee
The adiabatic index $\Gamma$ for the equations of state given in figure \ref{eos0} is shown in figure \ref{dps0}. Here, the adiabatic index $\Gamma$ 
is defined by
\be
\Gamma\equiv {\sigma_0+p_0\over p_0}{dp\over d\sigma}\,.
\ee

Since we obtain the nonrotating solutions of the thin shell gravastar given in figures \ref{sig0} and \ref{p0s0}, we may evaluate rotational effects 
on the structure of the spherical gravastars as small perturbations around them.  In this paper, we shall basically show numerical results of 
the perturbation quantities for $\epsilon=1$ or $\Omega=\Omega_k$. 

\subsection{ $\epsilon$-order rotational effects on the thin shell gravastar}
\begin{figure}[t]
\begin{center}
\includegraphics {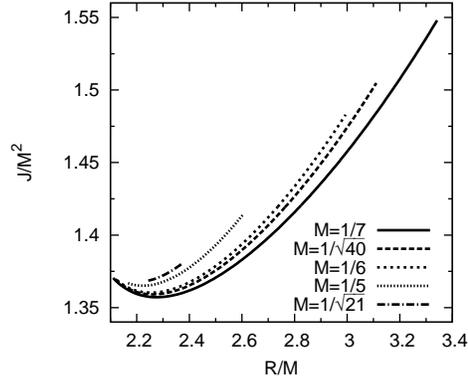}
\caption{The dimensionless angular momentum of the gravastar, $J/M^2$, as a function of $R/M$. }
\label{J}
\end{center}
\end{figure}
\begin{figure}[t]
\begin{center}
\includegraphics {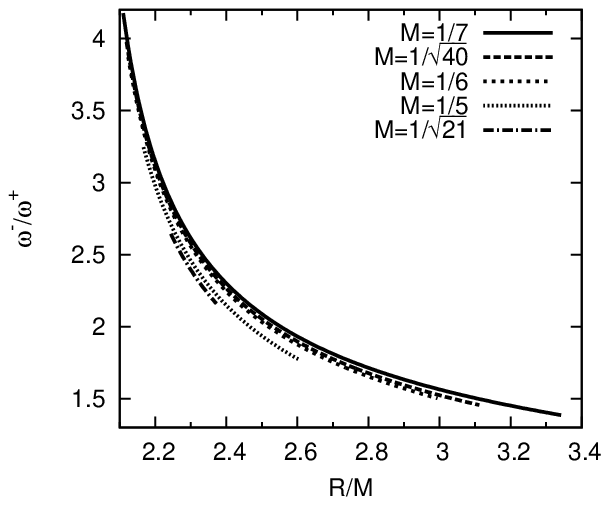}
\caption{The same as figure \ref{J} but for $\displaystyle{\omega^-\over\omega^+(R)}$. }
\label{win}
\end{center}
\end{figure}

The $\epsilon$-order rotational effects appear as the frame 
dragging effects in the exterior and interior spacetimes of the thin shell as discussed before. An observer at infinity  can measure the frame 
dragging effects in the exterior spacetime as the total angular momentum of the gravastar $J$. In figure \ref{J}, we show the dimensionless total 
angular momentum of the thin shell gravastar $J/M^2$ along the equilibrium sequences as a function of $R/M$. In this figure, we observe that 
the dimensionless angular momentum for the $M/L$ constant equilibrium sequences is basically an increasing function of $R/M$, however, it 
becomes a decreasing one for some equilibrium sequences when the gravastar is very compact (when $R/M \lesssim 2.25$). 
In figure \ref{win}, values of $\displaystyle{\omega^-\over\omega^+(R)}$ for the equilibrium sequences given in figures \ref{sig0} and \ref{p0s0} are given as functions of $R/M$. 
\subsection{$\epsilon^2$-order rotational effects on the thin shell gravastar} 
\begin{figure}[t]
\begin{center}
\includegraphics {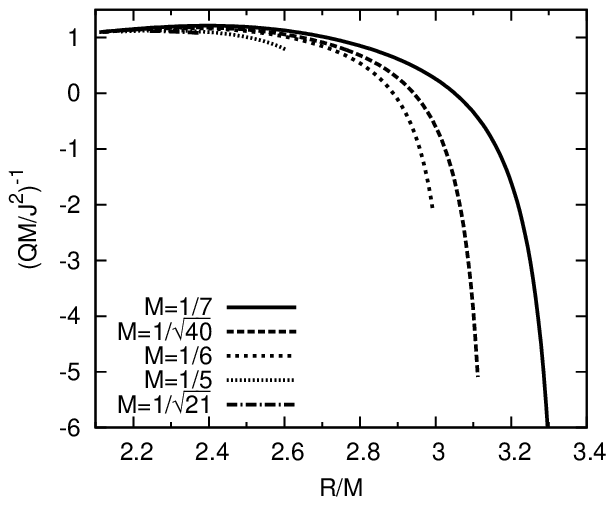}
\includegraphics {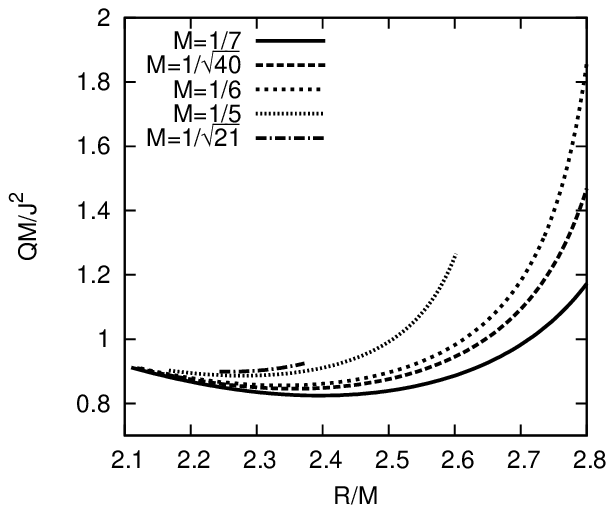}
\caption{Left:  the inverse of the dimensionless quadrupole moment, $(QM/J^2)^{-1}$,  as a function of $R/M$. Right:  the dimensionless quadrupole 
moment, $QM/J^2$, as a function of $R/M$ for the range of $2.1 \le R/M \le 2.8$.}
\label{Q-2}
\end{center}
\end{figure}
\begin{figure}[t]
\begin{center}
\includegraphics {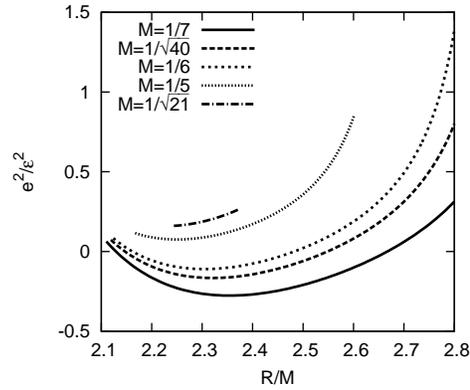}
\caption{The square of the ellipticity of the thin shell, $e^2/\epsilon^2$, as a function of $R/M$.}
\label{e-2}
\end{center}
\end{figure}
\begin{figure}[t]
\begin{center}
\includegraphics {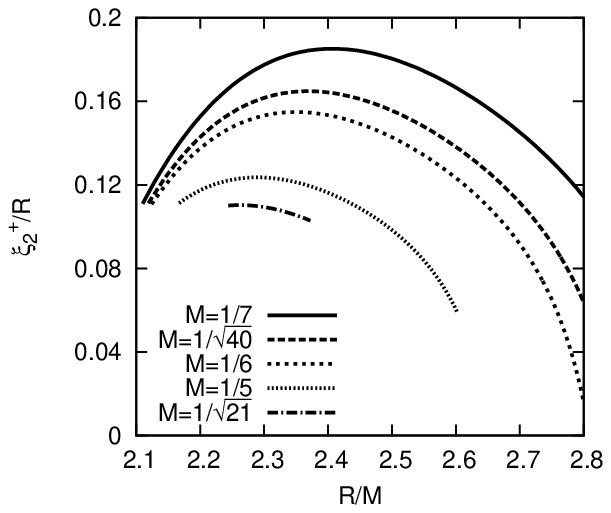}
\caption{The same as figure \ref{e-2} but for the normalized quadrupole component of the radial displacement of the thin shell, $\xi^+_2/R$. }
\label{z2-2}
\end{center}
\end{figure}
\begin{figure}[t]
\begin{center}
\includegraphics {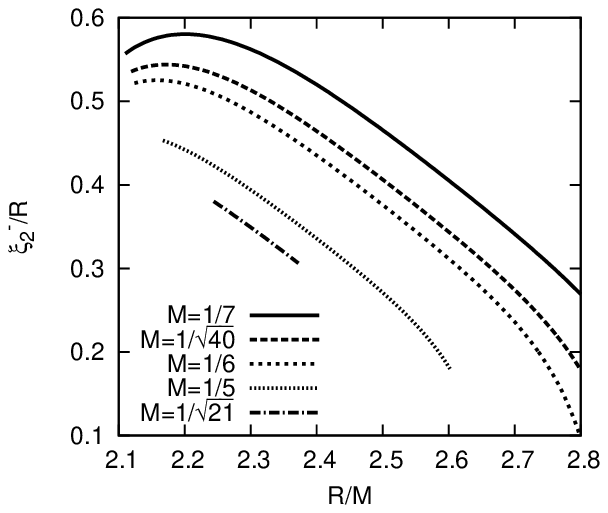}
\caption{The same as figure \ref{e-2} but for the normalized quadrupole component of the radial displacement  of the thin shell, $\xi^-_2/R$.}
\label{x2-2}
\end{center}
\end{figure}

First, we show physical quantities associated with quadrupole perturbations. Of these quantities, the quadruple moment is of the largest significance, 
because it represents information about the matter distribution that does not appear for spherically symmetric solutions and can be measured by 
analyzing motions of a test particle around it. As argued before, the slowly rotating Kerr black hole is characterized by the particular relation of 
the quadrupole moment $Q M/J^2 =1$. As a result, we may conclude that slowly rotating compact objects are not black holes if their quadrupole moment 
does not obey the relation $Q M/J^2 =1$. 

In figure \ref{Q-2}, we show the inverse of the dimensionless quadrupole moment $(QM/J^2)^{-1}$ and the dimensionless quadrupole moment $QM/J^2$ 
for the five equilibrium sequences characterized by a fixed value of $M$ (see section 4.1). In the left panel of this figure, we observe that values of 
$(QM/J^2)^{-1}$ vanish at a single point along the three equilibrium sequences with $M = 1/6$, $1/\sqrt{40}$, and $1/\sqrt{7}$. This means that because 
perturbation quantities diverge for these unperturbed solutions, the nonrotating thin shell gravastar solutions with $(QM/J^2)^{-1}=0$ cannot rotate with 
the perturbative approximation under the assumptions introduced in this study, i.e., assumptions of the perfect-fluid thin shell, (\ref{pful}), and 
the equation of state for the thin shell, (\ref{eos3}). The same singular behavior occurs in all the other quantities associated with quadrupole perturbations 
for the unperturbed solutions with $(QM/J^2)^{-1}=0$. Existence of these solutions with $(QM/J^2)^{-1}=0$ necessarily implies that the present perturbative 
approach also becomes irrelevant around such solutions. As seen from the left panel of figure \ref{Q-2}, however, nonrotating solutions with sufficiently lower and higher values of $R/M$ than those of the ones that 
cannot rotate with the perturbation approximation do not show any difficulty for investigation of rotational effects on their structure with the present treatment. 
Since we are interested in black hole like objects, hereafter, we will focus on the cases of the nonrotating models with $R/M \le 2.8$, for which no singular 
behavior appears in the perturbation solutions. 

\begin{figure}[t]
\begin{center}
\includegraphics {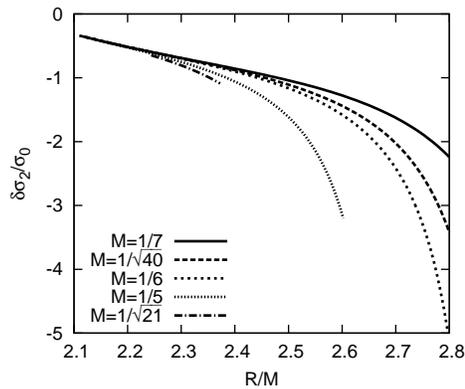}
\caption{The same as figure \ref{e-2} but for the normalized quadrupole component of the energy density perturbation of the thin shell, 
$\delta \sigma_2/\sigma_0$.}
\label{ds2-2}
\end{center}
\end{figure}

\begin{figure}[t]
\begin{center}
\includegraphics {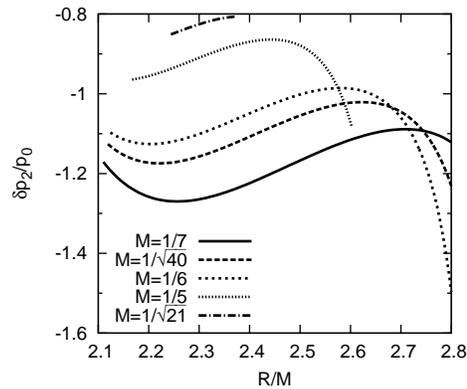}
\caption{The same as figure \ref{e-2} but for the normalized quadrupole component of the pressure perturbation of the thin shell, 
$\delta p_2/p_0$.}
\label{dp2-2}
\end{center}
\end{figure}

\begin{figure}[t]
\begin{center}
\includegraphics {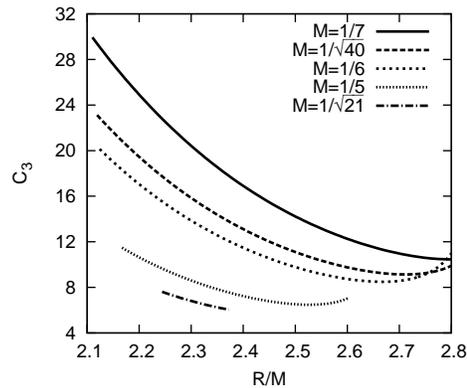}
\caption{The same as figure \ref{e-2} but for $C_2$.}
\label{C3-2}
\end{center}
\end{figure}

In the right panel of figure \ref{Q-2}, we show the dimensionless quadrupole moment $QM/J^2$ 
for the five equilibrium sequences characterized by a fixed value of $M$ and $R/M\le 2.8$.
From the right panel of figure \ref{Q-2}, we can see that some solutions have the same dimensionless quadrupole moment as that of slowly rotating Kerr black 
holes, i.e., $Q M/J^2 =1$. For these solutions, we cannot distinguish slowly rotating gravastars from slowly rotating black holes by using 
the quadrupole moment.  Another remarkable feature of the quadrupole moment of the slowly rotating thin shell gravastar is that for sufficiently 
compact models, values of $QM/J^2$ are less than unity, which are usually attributed to the matter distribution deformed prolately. For standard 
self-gravitating rotating objects, their matter distribution is oblately deformed due to the centrifugal force and their values of $QM/J^2$ are larger than unity. 
In order to see degree of the deformation of the thin shell associated with its shape, in figure \ref{e-2}, we show the normalized square of the ellipticity 
of the thin shell ${\rm e}^2/\epsilon^2$ 
for the five equilibrium sequences characterized by a fixed value of $M$, where ${\rm e}^2/\epsilon^2$ is defined by 
\be
{{\rm e}^2\over\epsilon^2}\equiv -3 \left( k_2^\pm (R)+ {\xi_2^\pm\over R} \right)\,. 
\ee
Note that we have $\displaystyle \left[\left[ k_2(R)+ {\xi_2\over R} \right]\right]=0$.
With this definition, positive (negative) values of ${\rm e}^2$ mean that slowly rotating thin shells are of oblate (prolate) shape. 
From figure \ref{e-2}, we see that for the case of $R/M \le 2.8$, the solutions belonging to the equilibrium sequences with $M=1/\sqrt{21}$ and $1/5$ 
always have positive values of the square of the ellipticity of the shell, while for the solutions  belonging to the other equilibrium sequences, 
values of ${\rm e}^2$ are basically negative but becomes positive at lower and higher values of $R/M$. Thus, some slowly rotating thin shell 
gravastars have an oblate shape thin shell but their values of $QM/J^2$ are less than unity.  
In figures \ref{z2-2} and \ref{x2-2}, we show the dimensionless radial displacements of the spherical shell $\xi_2^+/R$ and $\xi_2^-/R$, 
respectively. For the nonrotating gravastar models with $2.1 \le R/M \le 2.8$, we can see that  values of $\xi_2^+/R$ and $\xi_2^-/R$ are always positive.
The energy density and pressure perturbations rescaled by their unperturbed quantities, $\delta\sigma_2/\sigma_0$ and $\delta p_2/p_0$, 
are given as functions of $R/M$ in figures \ref{ds2-2} and \ref{dp2-2}, respectively. We observe that values of $\delta\sigma_2/\sigma_0$ and 
$\delta p_2/p_0$ are always negative in the range of $2.1 \le R/M \le 2.8$. 
Values of $C_3$ are shown in figure \ref{C3-2} along the five sequences of the nonrotating thin shell gravastars. 
From this figure, it is found that  $C_3 \le 30 $ for all the nonrotating solutions for $2.1 \le R/M \le 2.8$.
Since all the perturbation quantities given before are of the order of $O(1)$, we can conclude that our perturbation approach is suitable 
for the unperturbed solutions with $2.1 \le R/M \le 2.8$. 


\begin{figure}[t]
\begin{center}
\includegraphics {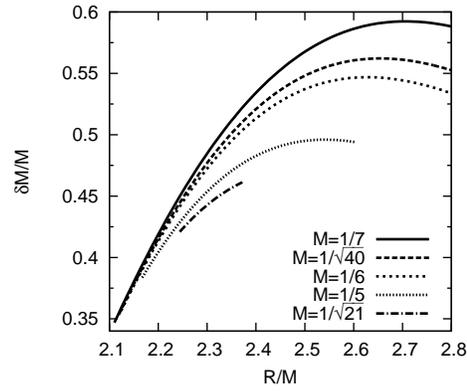}
\caption{The same as figure \ref{e-2} but for the relative change of the gravitational mass $\delta M/M$.}
\label{dM}
\end{center}
\end{figure}

\begin{figure}[t]
\begin{center}
\includegraphics {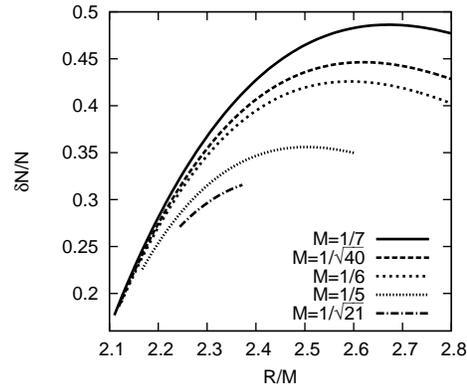}
\caption{The same as figure \ref{e-2} but for the relative change of the total particle number of the thin shell, $\delta N/N$.}
\label{dN}
\end{center}
\end{figure}

\begin{figure}[t]
\begin{center}
\includegraphics {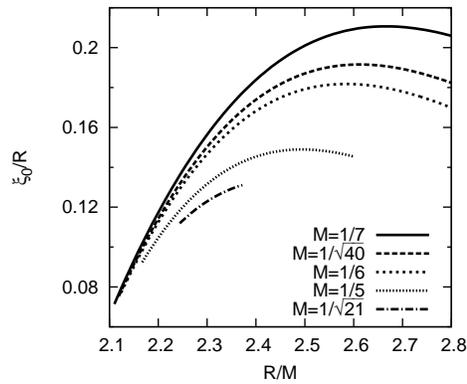}
\caption{The same as figure \ref{e-2} but for the normalized spherically symmetric component of the radial displacement of the thin shell $\xi_0/R$.}
\label{z0-2}
\end{center}
\end{figure}

\begin{figure}[t]
\begin{center}
\includegraphics {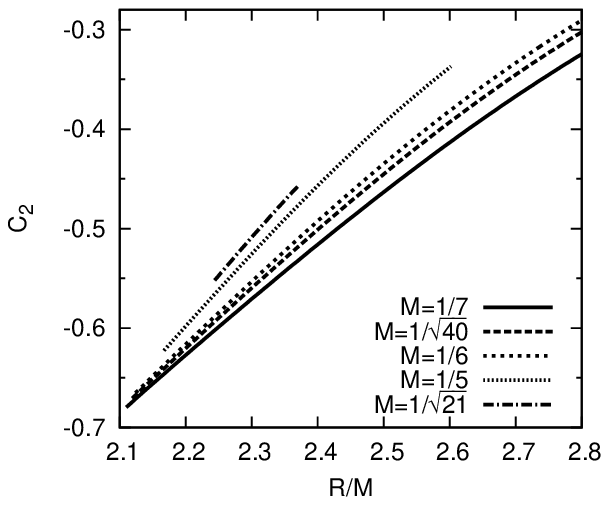}
\caption{The same as figure \ref{e-2} but for $C_2$.}
\label{C2-2}
\end{center}
\end{figure}

Next, we show the physical quantities associated with spherically symmetric perturbations. 
Since $\delta \sigma_0 = 0$ is assumed (see section 3.3) and values of $dp/d\sigma$ are finite as shown in figure \ref{eos0}, we have $\delta p_0 = 0$.  
In figures \ref{dM} and \ref{dN}, respectively, we show relative changes in the total gravitational mass, $\delta M/M$, and in the total particle number, $\delta N/N$, 
of the thin shell gravastar due to rotation for the five equilibrium sequences characterized by a fixed value of $M$.
In these figures, it is found that their behaviors are qualitatively similar to each other, that their values are always positive, and 
that there are the maximum values of $\delta M/M$ and $\delta N/N$ for the four equilibrium sequences with $M=1/7$, $1/\sqrt{40}$, $1/6$, and $1/5$.
The dimensionless radial displacement $\xi_0/R \equiv\xi_0^\pm/R$ is shown in figure \ref{z0-2}.
From this figure, we see that $\xi_0/R$'s as functions of $R/M$ show similar properties to those of $\delta M/M$ and $\delta N/N$. 
Values of $C_2$ as functions of $R/M$ along the five equilibrium sequences are shown in figure \ref{C2-2}.
In this figure, we see that values of $C_2$ are negative and of the order of $O(1)$.

\section{Discussions}

\subsection{Slowly rotating thin shells with isotropic pressure: limit of $L\rightarrow \infty$}
In section 3, we give master equations for obtaining slowly rotating thin shell gravastars. 
This formalism can also be applicable in the case of  $L\rightarrow \infty$. In this limit, 
the de Sitter core vanishes and the place occupied by a de Sitter core becomes vacuum. 
In other words, only a slowly rotating thin shell exists in a vacuum and the thin shell matter is 
the only source of the gravity. This is another interesting and fundamental situation apart from the gravastar. 
Therefore, similar but not the same situations have been so far considered by several authors, e.g., \cite{cohen,dlc,pf,pf2}. 

The quadrupole metric perturbations for the interior spacetime of the thin shell, given in (\ref{h2n}) and (\ref{k2n}), 
vanish in the limit of $L\rightarrow\infty$. However, this is not appropriate behavior for the metric perturbations inside 
the slowly rotating thin shell, because in general, the quadrupole metric perturbations need not vanish there. In order to obtain 
physically suitable solutions 
in this limit, we will use another integral constant $\widetilde{C}_3$, defined by $C_3=5 L^4 \, \widetilde{C}_3$. In the limit of 
$L\rightarrow\infty$, we then have appropriate solutions, given by $h_2 ^- = -k_2^-=\widetilde{C_3} r^2$. The integral constant 
$\widetilde{C_3}$ is determined with the same procedure described in the previous sections. For all the other perturbation 
quantities, we may take the $L \to \infty$ limit without any difficulty.

\begin{figure}[t]
\begin{center}
\includegraphics {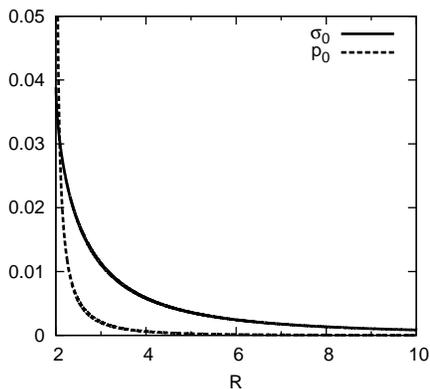}
\caption{Energy density $\sigma_0$ and pressure $p_0$ of the thin shell in vacuum asymptotically flat spacetime 
in the zero rotation limit as functions of $R/M$.}
\label{sig-p}
\end{center}
\end{figure}
\begin{figure}[t]
\begin{center}
\includegraphics {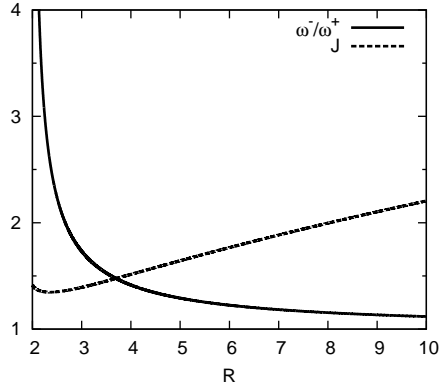}
\caption{The same as figure \ref{sig-p} but for the dimensionless angular momentum $J/M^2$ and $\omega^-/\omega^+(R)$. }
\label{J-f0}
\end{center}
\end{figure}
\begin{figure}[t]
\begin{center}
\includegraphics {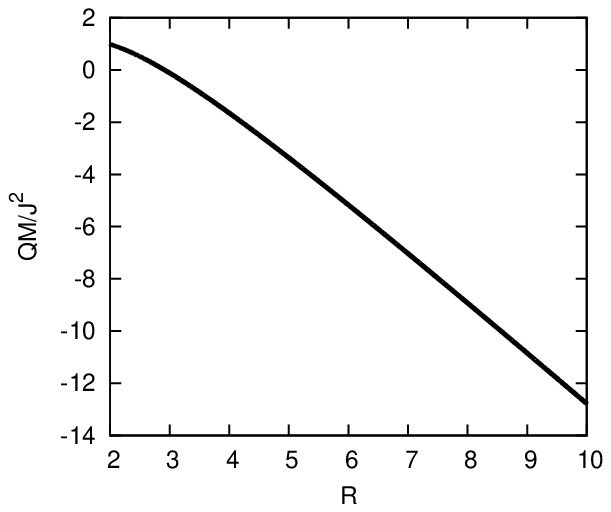}
\includegraphics {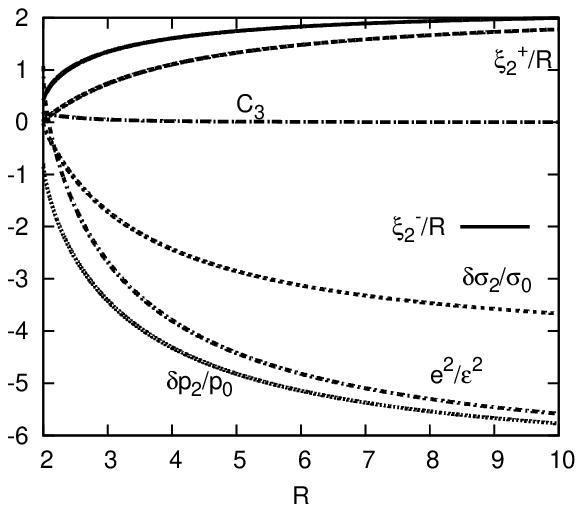}
\caption{Left: the dimensionless quadrupole moment of the thin shell, $Q M/J^2$, as a function of $R/M$. 
Right:  the quadrupole components of the perturbation quantities, $\xi^+_2/R$, $\xi^-_2/R$, $\delta p_2/p_0$, 
$\delta \sigma_2/\sigma_0$, $e^2/\epsilon^2$, and $\tilde{C}_3$ as functions of $R/M$.}
\label{P2-f2}
\end{center}
\end{figure}
\begin{figure}[t]
\begin{center}
\includegraphics {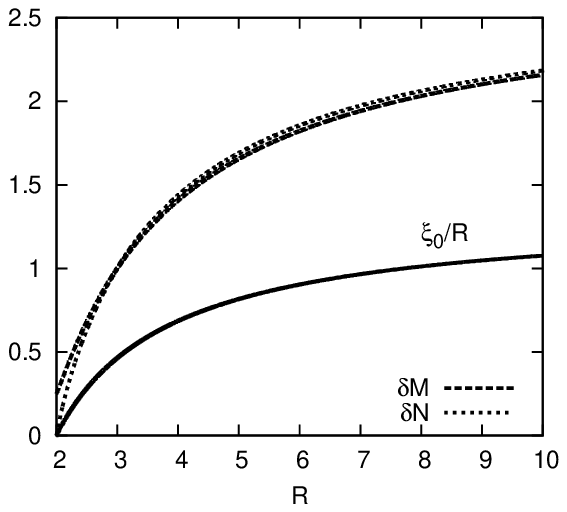}
\includegraphics {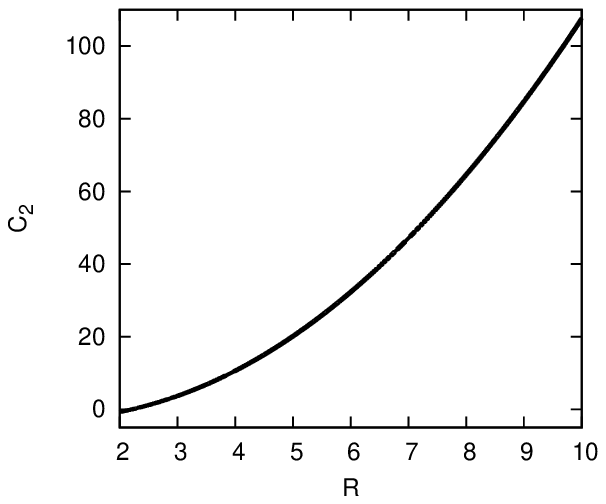}
\caption{Left:  the spherically symmetric components of the perturbation quantities, $\xi_0/R$, $\delta M/M$, and $\delta N/N$ 
as functions of $R/M$. Right:  $C_2$ as functions of $R/M$.}
\label{P0-f2}
\end{center}
\end{figure}
Since the limit of $L\rightarrow\infty$ is taken, it will be convenient to use the physical quantities nondimensionalized by a length scale of $M$. 
Henceforth, we therefore employ the units of $M=1$ to show the numerical results for the slowly rotating thin shell. For the weak energy 
condition, $\sigma_0>0$, to be satisfied, for the nonrotating thin shell, we require the inequality \ref{ineq0}. In the limit of $L\rightarrow\infty$, 
this condition becomes trivial and the thin shells with $\sigma_0>0$ are obtained for $R>2M$. In this paper, however, we focus on the cases of 
$R <10 M$ for the sake of conciseness. In figure \ref{sig-p}, the energy density and pressure of the nonrotating thin shell are shown as functions 
of the coordinate radius of the thin shell $R$. These solutions are unperturbed ones for investigating rotational effects on the structure of 
the spherical thin shell with the perturbation approach. From (\ref{sig00}) and (\ref{p00}), we may see that 
the dominant energy condition $\sigma_0\ge|p_0|$ for the nonrotating thin shell is met if $25 M/12 < R$ or $2.08 M \lesssim R$ is satisfied, 
which is consistent with results given in figure \ref{sig-p}.  In figure \ref{J-f0}, the angular momentum of the thin shell 
$J$ and the normalized angular velocity of the frame dragging inside the thin shell $\omega^-/\omega^+(R)$ are given as functions of $R/M$.
The basic properties of those quantities are similar to those of the gravastar, as can be confirmed in figure \ref{J}. 
The dimensionless angular momentum $J/M^2$ is basically an increasing function of the normalized radius of the shell $R/M$ but 
becomes a decreasing function for sufficiently compact thin shell models. 

The physical quantities associated with quadrupole perturbations are summarized in figure \ref{P2-f2}. The dimensionless quadrupole moment 
$QM/J^2$ and the other perturbation quantities are shown as functions of $R$ in the left and right panels of figure \ref{P2-f2}, respectively. 
In the left panel of this figure, we observe that the dimensionless quadrupole moment $Q M/J^2$ is always smaller than unity and becomes negative 
when the radius of the thin shell becomes sufficiently large, i.e., when $R \gtrsim 2.9M$. 
As can be seen in the right panel of figure \ref{P2-f2}, the normalized square of the ellipticity $e^2/\epsilon^2$ is basically negative, which means 
the shape of the slowly rotating thin shell is prolate. The dimensionless quadrupole radial displacement of the thin shell $\xi_2^{\pm}/R$ is always positive, 
while the quadrupole energy density and pressure perturbations are always negative. 

In figure \ref{P0-f2}, we show physical quantities of spherically symmetric perturbations $\delta M/M$, $\delta N/N$, $\xi_0/R$, (in the left panel) and 
$C_2$ (in the right panel) as functions of $R/M$. In this figure, we observe that the relative changes in the total gravitational mass $\delta M/M$ and 
the total particle number $\delta N/N$ show similar behavior each other and are always positive. The dimensionless radial displacement of the thin shell 
$\xi_0/R$ takes positive values regardless of the nonrotating thin shell models. 
In the right panel of figure \ref{P0-f2}, it is found that the integral constant $C_2$, which determines magnitude of metric perturbations inside the thin shell, 
increases rapidly as the radius of the thin shell $R/M$ increases. This means that with perturbation approaches, rotational effects on the structure of 
the thin shell cannot be investigated if the radius of the thin shell becomes sufficiently large because for such thin shells with sufficiently large radii, infinitesimally 
small angular velocity leads to a thin shell rotating with finite rotation velocity.  

For a slowly rotating thin shell with isotropic pressure, an interesting limiting case is the Newtonian limit, which is obtained in the limit of $M/R \to 0$. 
In the Newtonian limit, we have
\be
\sigma_0 \rightarrow {M\over 4\pi R^2} \,, 
\ee
\be
p_0 \rightarrow {1\over 4} \sigma_0 {M\over R} \,, 
\ee
\be 
J \rightarrow {2 \over 3} M R^2 \Omega_k \,, 
\ee
\beq
e^2 & \to -\f{20}{3} , \label{e_newton}\\
{QM\over J^2} & \to -{2 R \over M}. \label{q_newton}
\eeq
These results in the Newtonian limit are consistent with numerical results given in figures \ref{sig-p}--\ref{P2-f2}. 

With perturbation approaches, as mentioned before, self-gravitating slowly rotating thin shells were studied by several authors, e.g., \cite{cohen,dlc,pf,pf2}. 
De la Cruz and Israel~\cite{dlc} and Pfister and Braun~\cite{pf} obtained slowly rotating thin shells very similar to those discussed in this subsection. 
To investigate what is a possible source of the Kerr spacetime, de la Cruz and Israel obtained slowly rotating thin shells. Their main assumptions are 
the following: (i) the exterior spacetime of the thin shell is the Kerr one, (ii) the radial displacement in the exterior spacetime is given by 
$\xi^+ =k R(a/R)^2 (1-r_+/R)(1-r_-/R)\cos^2\Theta$, where $k$ is a constant, $R$ is the radius of the nonrotating thin shell, $a$ is the Kerr parameter, 
$r_+$ and $r_-$ are, respectively, the outer- and inner-horizon radii of the Kerr solution, and the radial coordinate $r$ is of the Boyer-Lindquist coordinates, 
(iii) there is no change 
in the total gravitational mass due to rotation ($\delta M =0$). Note that all the assumptions of de la Cruz and Israel's are different from ours. Their solutions 
are uniquely determined in terms of the total gravitational mass $M$, the Kerr parameter $a$, the radius of the thin shell $R$, and the constant $k$, which 
determines $\xi^+$. Thus, both the oblate and prolate shape thin shells are equally possible for their solutions. 
Making the assumptions: (i) uniformly rotating thin shells ($\Omega={\rm constant}$), (ii) the asymptotically flat spacetime, (iii) the interior spacetime 
of the thin shell is flat ($\widetilde{C}_3=0$), (iv) there is no change in the total gravitational mass due to rotation ($\delta M =0$), (v) the radial displacement 
is given by $\xi \propto \sin^2\Theta$, Pfister and Braun investigated rotational effects on the interior flat spacetime of the thin shell and on the thin shell structures. 
Note that Pfister and Braun's assumptions (iii)--(v) are different from ours. All the slowly rotating thin shells obtained by Pfister and Braun are of prolate shape. 
They also showed that even in the Newtonian limit ($M/R\rightarrow 0$), the shape of their slowly rotating thin shells is prolate. 
The common features of the slowly rotating thin shells obtained by De la Cruz and Israel~\cite{dlc} and Pfister and Braun~\cite{pf} are that 
the thin shell matter stress is described by anisotropic pressure. 

As for the slowly rotating thin shells obtained in the present study, as argued before, the shape of the thin shells is basically prolate except for 
the sufficiently compact thin shell cases. Like in the case of Pfister and Braun~\cite{pf}, the slowly rotating thin shells obtained in this study have the prolate shape 
even in the Newtonian limit [see figure \ref{P2-f2}, (\ref{e_newton}), and (\ref{q_newton})]. As shown so far, the slowly rotating self-gravitating thin 
shells frequently have prolate shape 
even in the Newtonian limit. Thus, we may guess that this occurrence of the slowly rotating thin shells with the prolate shape cannot be attributed 
to general relativistic effects like in the case of the centrifugal force reversal near the Schwarzschild black hole event horizon, see, e.g., \cite{abramo}. 
It is intuitively-plausible that in the Newton mechanics, self-gravitating objects become oblate as the rotation angular velocity increases 
if only the centrifugal force is additionally 
exerted on the matter. In the construction of the slowly rotating thin shell, however, we employ the matching of the two distinct spacetimes. 
In this means, the distribution of the infinitesimally thin matter is suitably determined for the two spacetimes to be glued continuously at 
a boundary hypersurface assumed particularly.  
Thus, it is not clear what is actually causing the occurrence of the prolate shape slowly rotating self-gravitating thin shells.

\subsection{Ergoregion of the slowly rotating thin shell gravastar within an accuracy up to the order of $\epsilon^2$}
\begin{figure}[t]
\includegraphics {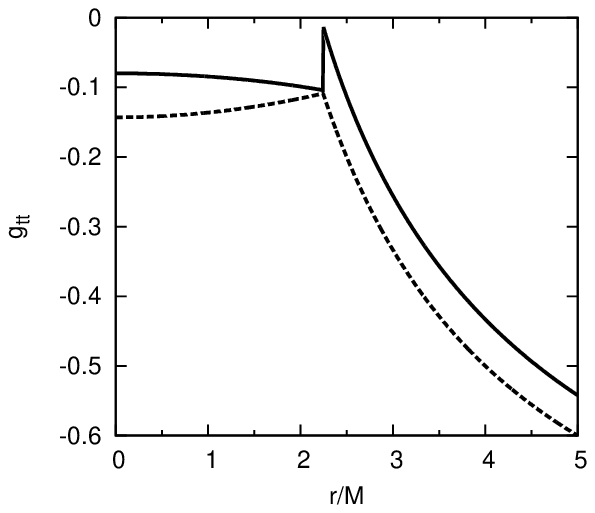}
\includegraphics {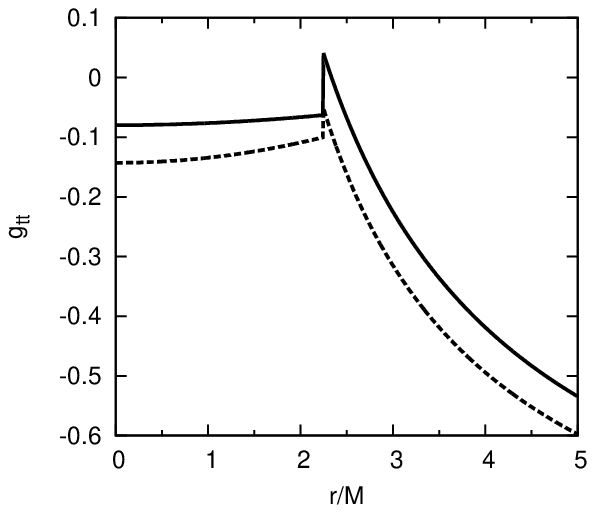}
\includegraphics {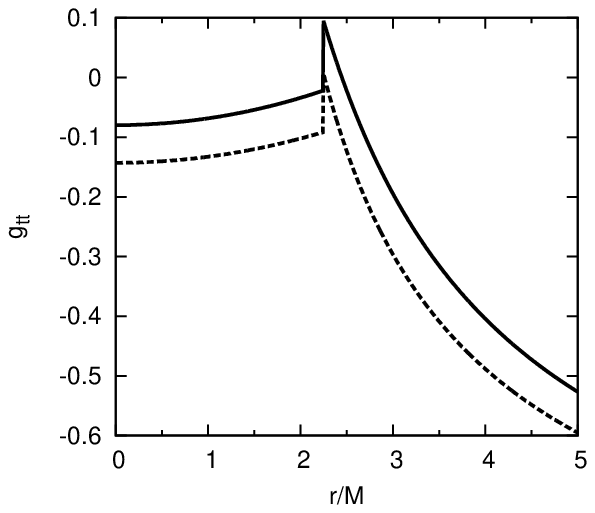}
\caption{$g_{tt}$ on the constant $\theta$ surfaces as functions of $r/M$ for the model with 
$M=1/\sqrt{21}$, $R=2.244M$, and $\epsilon^2 = 0.4$. Results for $\theta=0$, $\pi/4$, and $\pi/2$ are given in the left top, right, 
and left bottom panels, respectively. The solid curves correspond to the results given by the line element (\ref{in}) and the dashed curves to 
the results given by the line element (\ref{CS}).}
\label{gtt-no1}
\end{figure}
\begin{figure}[t]
\includegraphics {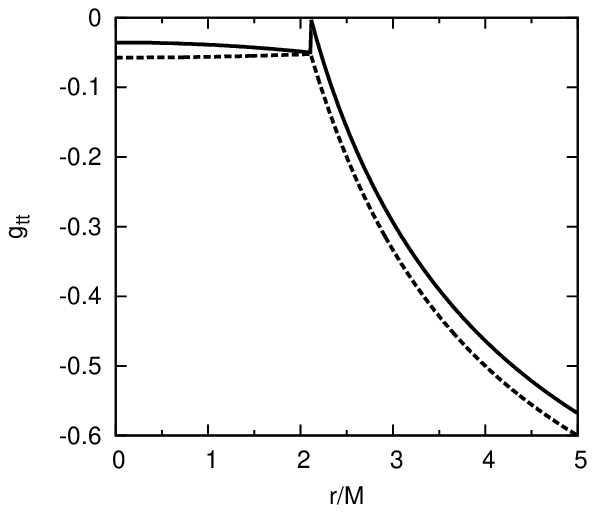}
\includegraphics {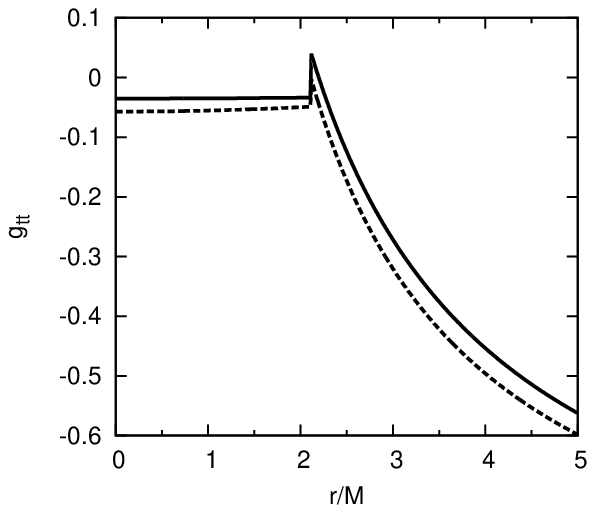}
\includegraphics {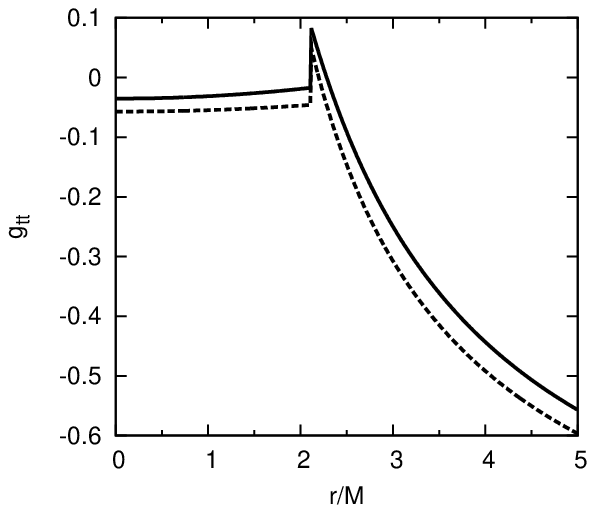}
\caption{The same as figure \ref{gtt-no1} but for the model  with $M=1/7$, $R=2.11M$, and $\epsilon^2 = 0.28$.}
\label{gtt-no2}
\end{figure}
In general relativity, sufficiently compact and rapidly rotating objects can in principle have the so-called ergoregions, in which no  
physical observer with $r={\rm constant}$, $\theta={\rm constant}$, and $\phi={\rm constant}$ can exist. Friedman has shown that an object having an ergoregion 
but no horizon in stationary asymptotic flat spacetime is unstable or marginally unstable to massless scalar and  electromagnetic 
perturbations~\cite{Friedman}. The same instability, of course, occurs for gravitational perturbations (see, e.g., \cite{vitor1}). Since it is secular and 
basically has very long growth time, the ergoregion instability can substantially operate only when it has sufficiently short growth time. 
Thus, calculations of the growth time for 
the ergoregion instability  in the system considered are required to see whether the ergoregion instability is driven effectively. 
However, general relativistic perturbations of fully stationary axisymmetric objects have in general been hard to deal with. Thus, 
a standard tractable way to treat ergoregions and ergoregion instability is to rely on an approximate method as follows: (i) Rotational effects 
are taken into account with a slow rotation approximation.  (ii) Results obtained for slowly rotating objects are then extrapolated 
to the case of rapid rotation. (iii) The extrapolation results are employed for investigation of properties of the ergoregions and 
ergoregion instability. Schutz and Comins  employed  this approximation method for the first time to study the ergoregion and 
ergoregion instability of rotating stars~\cite{Comins,Schutz}. The same method was applied for studies on the ergoregion instability 
of rotating gravastars~\cite{vitor1,cr2}. 

The ergoregion of stationary spacetime is defined by the region satisfying the inequality $g_{\mu\nu}t^\mu t^\nu > 0$ with $t^\mu$ 
being the time Killing vector. In the slow rotation approximation used in this study, this condition leads to
\be
g_{\mu\nu}t^\mu t^\nu=g_{tt}= -f(r)\left(1+2 \epsilon^2 h \right) + \epsilon^2\om^2\,r^2\sin^2 \theta > 0 \,. 
 \label{def_ergo}
\ee
If equation \ref{def_ergo} is satisfied for some values of $\epsilon$ satisfying $0< \epsilon \lesssim 1$, following Schutz and 
Comins~\cite{Comins,Schutz}, we approximately regard that there is the ergoregion in the spacetime considered. 
In their calculations, Schutz and Comins further neglect the term $2 \epsilon^2 h$ in \ref{def_ergo} from physical considerations, 
although it is not strictly a consistent approximation~\cite{Comins,Schutz}. In most studies on the ergoregion instability of 
rotating objects, this approximation method devised by Schutz and Comins, in which $h=0$ is assumed in (\ref{def_ergo}), have 
been widely used, see, e.g., \cite{vitor1,cr2,shin}. For the linear oscillation analysis, Schutz and Comins assumed that linear waves 
propagate through the spacetime described by the full metric, given by  
\be
ds ^2= -f(r) dt^2+g(r) dr^2 + r^2d\theta ^2+r^2\sin^2 \theta(d\phi -\epsilon\,\o dt)^2 \,, 
\label{CS}
\ee
where $f$ and $g$ are the metric functions in the non-rotating limit.  

In the spacetime described by the line element (\ref{in}), there is a useful particular frame that is called a locally non-rotating frame (LNRF) 
by Bardeen, Press, and Teukolsky~\cite{Bardeen}. Observers associated with the LNRF  follow trajectories defined by 
$d\phi=\epsilon\omega dt$, $dr=0$, and $d\theta=0$. Thus, along trajectories of the LNRF  observers, the line element is 
reduced to $ds_{\rm LNRF}^2=-f(r)\left\{1+2 \epsilon^2 (h_0(r)+h_2(r) P_2(\cos\theta))\right\} dt^2$. Just inside the black hole event horizon, 
$ds_{\rm LNRF}^2$ becomes positive. This means that  no physical observer with $dr=0$ and $d\theta=0$ is allowed inside the event horizon 
and that descent of physical observers along $r$ necessarily occurs there. Since we are concerned with objects with no event horizon nor 
spacetime singularity, in this study, we will consider the case of $ds_{\rm LNRF}^2<0$ only. 

In the approximate treatment 
by Schutz and Comins, we have $ds_{\rm LNRF}^2=-f(r)dt^2$. Therefore, $ds_{\rm LNRF}^2<0$ is automatically satisfied regardless of values of 
$\epsilon$ for objects with $f(r)>0$.  
When Schutz and Comins's approximate treatment is not adapted, or when the line element (\ref{in}) is assumed, 
$ds_{\rm LNRF}^2$ becomes positive for larger values of $\epsilon$ in some gravastar models considered in this study.  
We observe that  for such gravastar models, the region of $ds_{\rm LNRF}^2>0$ first appears on the symmetry axis as values of $\epsilon$ increase.  
Thus, the region of $ds_{\rm LNRF}^2>0$ appears if $h_0+h_2<0$, and the condition for occurrence of $ds_{\rm LNRF}^2>0$ is given by 
\be
\epsilon^2 > -\f{1}{2(h_0+h_2)} \quad {\rm if} \ h_0+h_2<0 \,. 
\label{ul}
\ee
From this inequality, we may obtain the condition for no appearance of the $ds_{\rm LNRF}^2>0$ region, given by 
\be
\epsilon^2 < \f{1}{{\rm Max}[-2(h_0+h_2)]} \equiv \epsilon_c^2 \quad {\rm if} \ h_0+h_2<0 \,, 
\label{def_ec}
\ee
where ${\rm Max}[q(r)]$ means the maximum value of the function $q(r)$. 
In this study, we require this condition for values of $ds_{\rm LNRF}^2$ to be negative everywhere. 
For the present gravastar models, we observe that the function $|h_0^+ +h_2^+|$ has the maximum value at $r=R$. 

\begin{figure}[t]
\begin{center}
\includegraphics {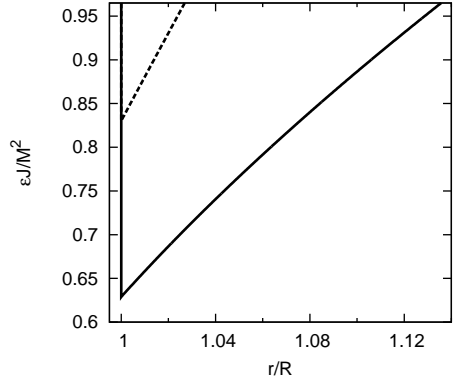}
\caption{Inner and outer radii of the ergoregion on the equatorial plane for the models with $M=1/\sqrt{21}$ and $R=2.244M$. 
The two intersection points between the solid or dashed curve and a $\epsilon J/M^2$ constant line are the radii of the ergoregion boundary. 
The solid curve corresponds to the result given by the line element (\ref{in}) and the dashed curve to 
the result given by the line element (\ref{CS}). }
\label{ergoregion-no1}
\end{center}
\end{figure}
\begin{figure}[t]
\begin{center}
\includegraphics {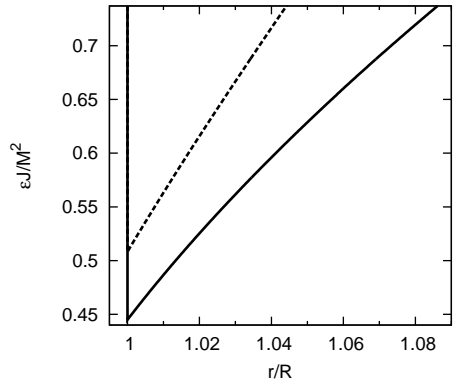}
\caption{The same as Fig.~\ref{ergoregion-no1} but for the model  with $M=1/7$, $R=2.11M$.}
\label{ergoregion-no2}
\end{center}
\end{figure}
To see properties of the ergoregion of the gravastar, the two models are examined: the models having $(M,R)=(1/\sqrt {21},2.244 M)$ 
and $(M,R)=(1/7,2.11M)$. The former is the most compact  model among the most massive equilibrium sequence (the $M= 1/\sqrt {21}$ 
equlinrium sequence), and the latter is the most compact model among the gravastar models calculated in this study. 
In figures \ref{gtt-no1} and \ref{gtt-no2}, fixing a value of $\theta$, we exhibit $g_{tt}$'s as functions of $r/M$ for the gravastars 
with $(M,R)=(1/\sqrt {21},2.244 M)$ and $(M,R)=(1/7,2.11M)$, respectively. For the models given in figures \ref{gtt-no1} and \ref{gtt-no2}, 
the values of $\epsilon^2$ are assumed to be $\epsilon^2 = 0.4<\epsilon_c^2 \approx 0.449$ and 
$\epsilon^2 = 0.28<\epsilon_c^2 \approx 0.289$, respectively. In these figures, the solid and dashed curves correspond to the results given 
by the line elements \eref{in} and \eref{CS}, respectively. The left top, right and left bottom panels in figures \ref{gtt-no1} and \ref{gtt-no2} show 
the results for $\theta=0$, $\pi/4$, and $\pi/2$, respectively. Note that in these figures, $g_{tt}$'s for $r<R$ are different coordinate components 
from those for $r>R$, because the coordinate functions inside the thin shell are defined independently from the ones outside the thin shell, i.e., 
$g_{tt}=g_{t^- t^-}$ for $r<R$ and $g_{tt}=g_{t^+ t^+}$ for $r>R$. 
Thus, jumps in $g_{tt}$'s appearing at $r=R$ are of spurious structure and have no physical meaning. 
We observe in these figures that  values of $g_{tt}$ decrease everywhere when the $2\epsilon^2 h $ term in (\ref{in}) is  neglected. 
This implies that for the gravastar considered in this study, use of the line element (\ref{CS}) results in a smaller ergoregion. 
From figures \ref{gtt-no1} and \ref{gtt-no2}, it is found that there is no ergoregion inside the thin shell. We also see from these figures that values 
of $g_{tt}$ are negative along the symmetry axis (i.e., there is no ergoregion on the symmetry axis) and that  the cross-section of the ergoregion perpendicular to 
the symmetry axis is the maximum on the equatorial plane like in standard rotating objects. 
In figures \ref{ergoregion-no1} and \ref{ergoregion-no2}, we show the boundaries of ergoregion on the equatorial plane 
for the models having $(M,R)=(1/\sqrt {21},2.244 M)$ and $(M,R)=(1/7,2.11M)$, respectively. In these figures, again, 
the solid and dashed curves correspond to the results given by the line elements~(\ref{in}) and (\ref{CS}), respectively. 
The two intersection points between the solid or dashed curve and a $\epsilon J/M^2$ constant line are the radii of 
the ergoregion boundary for the gravastar rotating at a given value of $\epsilon J/M^2$. The minimum value of $\epsilon J/M^2$, 
$\epsilon_{\rm min} J/M^2$, for each curve indicates the value of the dimensionless angular momentum at which the ergoregion 
first occurs. The maximum values of $\epsilon J/M^2$ in figures \ref{ergoregion-no1} and \ref{ergoregion-no2} are given by 
values  of $\epsilon_c J/M^2$ determined by \eref{def_ec}. 
Note that ${g}_{t^- t^-}$'s are always negative for the models considered in this study, so that the inner boundary of the ergoregion 
coincides with a radius of the thin shell. From figures \ref{ergoregion-no1} and \ref{ergoregion-no2}, we see that the ergoregion forms at 
a smaller value of $\epsilon J/M^2$ when the line element~(\ref{in}) is assumed. The parcentage differences between $\epsilon_{\rm min} J/M^2$'s  
given by the line elements~(\ref{in}) and (\ref{CS}) are approximately $24.4\%$ and $12.8\%$ for the models with 
$(M,R)=(1/\sqrt {21},2.244 M)$ and $(M,R)=(1/7,2.11M)$, respectively. These relative differences become smaller as 
the gravastar model becomes more compact as pointed out by Schutz and Comins~\cite{Schutz}. 

\section{Conclusions}
We have constructed the solutions of slowly rotating gravastars with a thin shell. In the zero-rotation limit, we consider the gravastar 
composed of a de Sitter core, a thin shell, and Schwarzschild exterior spacetime. The rotational effects are treated as small axisymmetric and stationary 
perturbations. The perturbed internal and external spacetimes are matched with a uniformly rotating thin shell. We assume that the angular velocity of 
the thin shell, $\Omega$, is much smaller than the Keplerian frequency of the nonrotating gravastar, $\Omega_k$. The solutions within an accuracy up to 
the second order of $\Omega/\Omega_k$ are obtained. The thin shell matter is assumed to be described by a perfect fluid and to satisfy the dominant 
energy condition in the zero-rotation limit. In this study, we assume that the equation of state of the thin shell matter is given by 
$dp/d\sigma = (\partial p_0/\partial R)/(\partial \sigma_0 /\partial R)$. In other words, the equation of state for perturbations is the same 
as that of the unperturbed solution. The spherically symmetric component of the energy density perturbations, 
$\delta\sigma_0$, is assumed to vanish independently of the rotation rate, i.e., $\delta \sigma_0 = 0$ is always assumed.  

Using the nonrotating thin shell gravastar stable against radial perturbations derived by Visser and Wiltshire~\cite{visser} as unperturbed solutions, 
under the assumptions mentioned before, we numerically evaluate the rotational corrections to the structure of the gravastar up to the second order of 
$\Omega/\Omega_k$. For nonrotating gravastar models, we choose five sequences of equilibrium solutions characterized by fixed values of 
$M/L=1/7$,~$1/\sqrt{40}$,~$1/6$,~$1/5$,~and~$1/\sqrt{21}$.  By changing  values of $R/M$, for a given value of $M/L$, we may obtain one sequence 
of the non-rotating gravastar. The maximum and minimum values of $R/M$ are determined for the thin shell to satisfy the dominant energy condition. 
For each sequence of the nonrotating gravastar models with $M \le 1/6$, there is a singular non-rotating gravastar model in the sense that 
the master equations for the quadrupole perturbations have no solution. In other words, the gravastar cannot rotate perturbatively for such singular 
non-rotating gravastar models. These singular solutions have a value of $R/M$ in the range of $2. 9 \lesssim R/M \lesssim 3$. To avoid the occurrence of 
such singular solutions, in this study, we focus on the nonrotating gravastar models with $R \le 2.8 M$ and examine rotational effects on the structure 
of these gravastars.  
Note that some authors have shown that horizon-less compact objects with stable photon circular orbits are nonlinearly unstable \cite{jk,vitor3}, 
which suggests that the gravstars with $R \le 2.8 M$ are nonlinearly unstable. However, we think that because of the nonlinearity, it is required further 
investigations to say that this nonlinear instability is general, although it indeed operates for some situations. 
Interesting findings in this study are the following: (i) Some rotating solutions have the dimensionless quadrupole moment, $QM/J^2$, the same as 
that of the Kerr black hole, i.e., $QM/J^2=1$. This means that if those gravastars are incidentally observed, we cannot distinguish gravastars from 
black holes by using their values of the mass, $M$, spin angular momentum, $J$, and mass quadrupole moment, $Q$. (ii) Most gravastar models 
obtained in this study satisfy the relation of $QM/J^2 < 1$. This is absolutely different from that of the standard rotating compact stars like neutron 
stars, for which $QM/J^2 > 1$ is satisfied. (iii) Some rotating gravastar models are prolate. This is again an opposite behavior to that in the standard 
rotating compact stars. The reasons for occurrence of (ii) and (iii) are not yet clear. 

The  present slowly rotating gravastars become slowly rotating thin shells with isotropic pressure in vacuum asymptotically flat spacetime in 
the limit of the zero cosmological constant (in the limit of $L\to\infty$). Taking the limit of $L\to\infty$, thus, we obtain the slowly rotating thin 
shell models and investigate their properties. For the thin shell, the dominant energy condition is satisfied if 
$R/M > 25/12\approx 2.08$. Thus, we may take the Newton limit of the slowly rotating thin shell in vacuum asymptotically flat spacetime.  
We find that many rotating thin shells calculated in this study are prolate and that dimensionless quadrupole moment for most of them satisfy $QM/J^2<1$ and 
become negative when the radius of the shell is larger than $2.9M$. This means that under our assumptions, slowly rotating thin shells are prolate 
even in the Newton limit. Thus, we do not think that the appearance of the prolate-shape slowly rotating gravastar is attributed to the 
general relativistic effects. 

We also examine properties of the ergoregion for the present thin shell gravastar models since its existence results in secular instability and 
their rotational energy can be extracted through the ergoregion instability. To examine the ergoreion of the thin shell gravastar, we basically 
follow the prescription devised by Schutz and Comins~\cite{Schutz,Comins}, in which the $\epsilon^2$-order rotational effects are partly 
included in the spacetime for the sake of tractability. However, we also consider the case where the $\epsilon^2$-order rotational effects are 
fully included in the spacetime. We find the following properties: (i) The ergoregion forms only outside the thin shell. (ii) Percentage 
differences between the minimum values of $\epsilon J/M^2$ at which the ergoregion first forms obtained with the Schutz and 
Comins's treatment and by the spacetime including all the $\epsilon^2$-order rotational effects are  $\approx 20\%$.  
(iii) This relative difference becomes smaller for more compact gravstar models as expected by Schutz and Comins~\cite{Schutz,Comins}. 
(iv) The ergoregion determined with the Schutz and Comins's treatment is smaller than that determined by the spacetime including all 
the $\epsilon^2$-order rotational effects.

\ack
N.U. acknowledges financial support provided under the European Union's FP7 ERC Starting Grant ``The dynamics of black holes:
testing the limits of Einstein's theory'' Grant Agreement No. DyBHo--256667. 
N.U. thanks Toshifumi Futamase for his kind hospitality at Tohoku University. 
S.Y. thanks Jos{\'e} Lemos and Vitor Cardoso for their kind hospitality at Institute Superior T{\'e}cnico, 
where some parts of this work were done. This work was supported in part by a Grant-in-Aid
for Scientific Research from JSPS (Grant No. 24540245). 

\appendix
\section{Extrinsic curvature}
The extrinsic curvature is approximated by $K^a_b={^{(0)}{K}}^a_b+\epsilon \, {^{(1)}{K}}^a_b+\epsilon^2 \, {^{(2)}{K}}^a_b+O(\epsilon^3)$
for slow rotation approximation. The explicit forms of the nonzero components of ${}^{(2)}{K}^a_b$ are given as follows: 
\beq
^{(2)}{K}^T_T &=  \f{1}{ R \sqrt{f^+}^3 }   \Big[ 2m_0 f^{\prime} -4R f^2 h_0^{\prime} +R\xi_0\left ((f^{\prime})^2-\f{4f(f-1)}{R^2} \right )+\f{4R^3}{3} \om \om^{\prime}\nn \\
 & +\Big \{2m_2 f^{\prime}-3 R f^2 h_2^{\prime} \left. \left . +R\xi_2\left ((f^{\prime})^2-\f{4f(f-1)}{R^2} \right )-\f{4R^3}{3} \om \om^{\prime} \right \}P_2  \right ] , \nn \\
 &
\eeq
\beq
^{(2)} {K}^{\Theta}_{\Theta} &+^{(2)}{K}^{\Phi}_{\Phi} \nn \\
& = \f {1} { 3R^2 \sqrt {f}} \Big [ 6m_0+3\xi_0(2 f -Rf^{\prime}) -R^4 \om \om^{\prime}  \nn \\ 
&- \Big \{6m_2 +3 \xi_2 (2 f -Rf^{\prime}-6 ) -6 R^2 f k_2 ^{\prime} + R^4 \om \om^{\prime} \Big \} P_2 \Big ] ,
\eeq
\beq
^{(2)} {K}^{\Theta}_{\Theta} &- ^{(2)}{K}^{\Phi}_{\Phi}  = \f {6\xi_2+R^4 \om \om^{\prime}} { 2 R^2 \sqrt {f}} \sin^2 \Theta, 
\eeq
\beq
 ^{(2)}K &= \f{1}{R^2 \sqrt{f}^3 } \Big [ 2m_0(4f+Rf^{\prime})-4R^2 f^2 h_0^{\prime}+\xi_0 \left(4f(1+f-Rf^{\prime})+R^2 (f^{\prime})^2 \right ) \nn \\
  &+ \Big \{ 2 m_2 (4f+R f^{\prime})-4 R^2 f^2 (h_2^{\prime}+2k_2^{\prime})   +\xi_2 \left (4 f (f-R f^{\prime} -5)+R^2 (f^{\prime})^2 \right) \Big \} P_2 \Big ].\nn \\
  & 
\eeq
\section{The $\epsilon^2$-order perturbations of the energy density and pressure }
%
The quadrupole components of the energy density and pressure perturbations are, respectively, given by 
\beq
\delta \sigma_2 & = \f{1}{4 \pi R^2} \left \{ \f{2 \xi^-_2 -m_2^-}{\sqrt{f^-} }-\f{(2 J -R^3 \Omega_k)^2}{3 R^3 f^+\sqrt{f^-}}  - R^2 [[\sqrt{f} k_2^{\prime}]] -\f{3M+2R}{R\sqrt{f^+}}\xi_2^+ \right. \nn\\
& +\f{J(-2 J+R^3 \Omega_k)+R^3 m_2}{\sqrt{f^+}}   \left. +\f{(R^3 \Omega_k -2 J)(J +R^2 (R-3 M)\Omega_k)}{3 R^3 \sqrt{f^-}^3 }   \right \},\nn \\
&
\eeq
\beq
\delta p_2 & = \f{1}{8\pi R^2} \left \{R^2 \left [ \left[ \sqrt{ f} (h_2^{\prime} + k_2 ^{\prime})  \right ] \right ] \right. \nn \\
& +\f{1}{R\sqrt{f^+}^3} \left ( \f {(2 R^2 -3 M R +3 M^2) \xi^+_2 }{R}-(R-M) m_2 ^+ \right ) \nn \\
& + \f{R^3 \Omega_k-2 J}{3 R\sqrt{f^+}^3} \left( \f{ 2 J(2 R-3 M)}{R^3} + (R-3M) \Omega_k) \right) \nn \\
& \left .-\f {(R^3 \Omega_k -2 J) ^2} {3R^3 f^+ \sqrt{f^-}}+\f {(L^2 -2 R^2) m_2^--(2 L^2 -3 R^2) \xi^-_2} {L^2 \sqrt{f^-} ^3} \right \} .
\eeq
The spherically symmetric components of the energy density and pressure perturbations are, respectively, given by 
\beq
\delta \sigma_0 & = \f{1}{4 \pi R^5} \left \{  \f{( \Omega_k -2 J)^2}{3  f^+ \sqrt{f^-}}\right .+ \f{1}{ \sqrt{f^+}} \Big ( J (2 J-R^3 \Omega_k)  +  m_0^+ \nn \\
& \left . \left. +\f{(2 J -R^3 \Omega_k)(J+R^4 (R-3M)\Omega_k)}{3 f^+} \right ) \right \}-\f{2(\sigma_0 +p_0)}{R}\xi_0,
\eeq
\beq
\delta p_0 & = -\f {1} {8 \pi R^2} \left \{\f{(R-M)m_0}{R \sqrt{f^+}^3} -R^2\sqrt {f^+}( h_0^+)^{\prime}- \f {(2 J - R^3 \Omega_k)^2} {3 R^3 f^+ \sqrt {f^-}}  \right. \nn \\
& \left.+ \f {(R^3 \Omega_k -2 J) (2 J (2 R-3 M)+R^2 (R-3M) \Omega_k)} {3 \sqrt{f^+} R^3} \right \}  +\f{\partial p_0}{\partial R}\xi_0.
\eeq
%
\section{The explicit form of $B$}
%
The explicit form of $B$ is given by 
\beq
B & = \left [ \f{U }{8 R f^+ \sqrt{f^-}} \left (\f{\sqrt{f^+}}{\sqrt{f^-}}+\f{3M -R}{R} \right )\right .  +\f{J^2 (M+R)}{8 M R^3 } \left (\f{1}{\sqrt{f^-}} -\f{R-M}{R\sqrt{f^+}} \right ) \nn \\
& + \f{1}{24R^4 (f^+)^2} \left ( \f{M \sqrt{f^+}(2J -R^3 \Omega_k)^2}{f^-} -\f{2J^2(M-R)(6M^2-R^2)}{\sqrt{f^+}R^2} \right. \nn\\
& +\f{(2J -R^3 \Omega_k)(2J(3M-R)-R^2(6M^2-R^2)\Omega_k)}{\sqrt{f^-}}\nn\\
& \left. +\f{R^2\Omega_k (R^2-M R -3 M^2)(2J+R^2 (R-3M)\Omega_k)}{\sqrt{f^+}} \right ) \nn \\
& +\f{dp}{d \sigma} \left \{U \left (-\f{2 R+3M}{4 R \sqrt {f^-}} +\f{L  \sqrt{f^-} G}{2 F} \right.  -\f{L M W}{R\sqrt{f^+} F} \right ) \nn \\
& - \f{M+R}{4 R f^+} \left (\f{(2 J -R^3 \Omega_k)^2}{R^3 \sqrt{f^-}} -\f{\Omega_k (2J+R^2 (R+3M) \Omega_k)}{\sqrt{f^+}}\right ) \nn \\
& +\f{J^2 (5M - 4 R)}{4 R^3 M \sqrt{f^+}^3}+\f{L W}{3 R^4(f^+)^2F} \left (M \sqrt{f^+} (2 J -R^3 \Omega_k)^2\right. \nn \\
&\left. -  \sqrt{f^-} \left (\f{J^2 (-8M^2 + 9M R -3 R^2)}{M} \right.    +M R^3 \Omega_k (2J+R^2 (R-3M) \Omega_k) \Big ) \bigg ) \bigg \} \right] \nn \\
& \times \left [\f{1}{8}\left (\f{R-3M}{R\sqrt{f^+}} -\f{1}{\sqrt{f^-}} \right ) \left ( \f{V M}{R\sqrt{f^+} \sqrt{f^-}}+R Q^2_2(X) \right ) \right. \nn \\
& + \f{dp}{d \sigma} \left \{ V \left (\f{3M+2R}{4 R \sqrt{f^-}} -\f{L \sqrt{f^-}G}{2F}-\f{LMW}{R \sqrt{f^+} F} \right )\right. -\f{R L \sqrt{f^-}W}{F } Q^2_2(X) \nn \\
& \left. \left.  - \f{\sqrt{f^+}}{4 M} \left ( \f{3 R^2 (R+2M)}{2M} \log Y+\f{6 M^3 + 8 M^2 R -3 M R^2 -3 R^3}{R f^+} \right ) \right \} \right ] ^{-1},
\eeq
where
\beq
X & = \f{R}{M}-1, \\
Y &=  \f{R}{R-2M} , \\
U & = \f{1}{R^3}\left \{\f{R^3 \Omega_k -2 J}{3 f^+} \left(1 +\f{M Lf^-G}{R f^+F} \right ) \right. \nn \\
& \times \left(\f{R^3\Omega_k -2 J}{\sqrt{f^-}}  -\f{2J(2R-3M) +R^3 (R-3M)\Omega_k}{R\sqrt{f^+}} \right )  \nn \\
& \left. -\f{J^2}{M\sqrt{f^+}} \left (R+2M-\f{Lf^-G(M+R)}{ F  } \right )\right \} \nn \\
& \times \left \{ \f{1}{\sqrt{f^+}} \left(1+\f{R^2G}{LF} \right )-\f{1}{\sqrt{f^-}} \left (1+\f{MLf^-G}{Rf^+ F} \right ) \right \}^{-1}, \\
V & = \left \{ \f{Lf^-G}{FM} \left (-\f{2M^2(M+ 2 R)+ 3 R^2(R-3M)}{R^2 f^+}  +\f{3R^2f^+}{2M} \log Y\right ) \right.\nn \\
&-\f{2M^2-3M R -3 R^2}{M R}  \left.- \f{3 R (R^2-M^2)log Y}{2M^2} \right \}  \nn\\
& \times \left \{ \f{1}{R}+\f{RG}{LF} -\f{\sqrt{f^+}}{R\sqrt{f^-}} \left (1+\f{MLf^-G}{Rf^+ F} \right ) \right \}^{-1},\\
W & = -3 L^2 R+2 R^3 +3 L^3 f^-\mbox{Arctanh}(R/L).
\eeq

\vspace{-0.2cm}
\section*{References}


\begin{thebibliography}{99}

\bibitem{mar} Markov M A 1982 Limiting density of matter as a universal low of nature {\it Sov. Phys. JETP Lett.}  {\bf 36} 265 
%
\bibitem{gravastar} Mazur P O and Mottola E 2001 Gravitational condensate stars {\it Preprint} gr-qc/0109035 
\nonum Mazur P O and Mottola E 2004 Gravitational vacuum condensate stars {\it Proc. Natl. Acad. Sci.} {\bf 111} 9545
%
\bibitem{visser} Visser M and Wiltshire D L 2004 Stable gravastars -- an alternative to black holes? {\it Class. Quantum Grav.} {\bf 21} 1135
%
\bibitem{carter} Carter B M N 2005 Stable gravastars with generalized exteriors {\it Class. Quantum Grav.} {\bf 22} 4551 

\bibitem{cattoen} Cattoen C, Faber T and Visser M 2005 Gravastars must have anisotropic pressures {\it Class. Quantum Grav.} {\bf 22} 4189 

\bibitem{cr} Chirenti C B M H and Rezzolla L 2007 How to tell a gravastar from a black hole {\it Class. Quantum Grav.} {\bf 24} 4191 
%
\bibitem{vitor1} Cardoso V, Pani P, Cadoni M and Cavaglia M 2008 Ergoregion instability of ultracompact astrophysical objects {\it Phys. Rev.} D {\bf 77} 124044

\bibitem{pani}  Pani P, Berti E, Cardoso V, Chen Y and Norte R 2009 Gravitational wave signatures of the absence of an event horizon: Nonradial oscillations of a thin-shell gravastar {\it Phys. Rev.} D {\bf 80} 124047 

\bibitem{Ryan} Ryan F D 1995 Gravitational waves from the inspiral of a compact object into a massive, axisymmetric body with arbitrary multipole moments {\it Phys. Rev.} D {\bf 52} 5707 

\bibitem{cr2} Chirenti C B M H and Rezzolla L 2008 Ergoregion instability in rotating gravastars {\it Phys. Rev.} D {\bf 78} 084011 

\bibitem{uchi} Uchikata N and Yoshida S 2014 Slowly rotating regular black holes with a charged thin shell {\it Phys. Rev.} D {\bf 90} 064042 

\bibitem{hartle} Hartle J B 1967 Slowly Rotating Relativistic Stars. I. Equations of Structure {\it Astrophys. J. }{\bf 150} 1005  
\nonum Hartle J B and Thorne K S 1968 Slowly rotating relativistic stars. II. Models for neutron stars and supermassive stars {\it Astrophys. J. }{\bf 153} 807

\bibitem{thorne} Thorne K S 1971  General relativity and cosmology {\it edited by Sachs R K (Academic Press, New York)}

\bibitem{chandra} Chandrasekhar S and Miller J C 1974 On slowly rotating homogeneous masses in general relativity {\it Mon. Not. R. Astron. Soc.} {\bf 167} 63 

\bibitem{israel} Israel W 1966 Singular hypersurfeces and thin shells in general relativity {\it Nuovo Cimento} {\bf 44B} 1 

\bibitem{ba}  Barrab\'es C and Israel W 1991 Thin shells in general relativity and cosmology: The lightlike limit {\it Phys. Rev.} D {\bf 43} 1129 


\bibitem{jk} Keir J 2014 Slowly Decaying Waves on Spherically Symmetric Spacetimes and an Instability of Ultracompact Neutron Stars arXiv:1404.7036 
\bibitem{vitor3} Cardoso V, Crispino L C B, Macedo C F B, Okawa H and  Pani P 2014 Light rings as observational evidence for event horizons: Long-lived modes, ergoregions and nonlinear instabilities of ultracompact objects {\it Phys. Rev.} D {\bf 90} 044069 

\bibitem{cohen} Brill D R and Cohen J M 1966 Rotating Masses and Their Effect on Inertial Frames {\it Phys. Rev.} {\bf 143} 1011 
\nonum Cohen J M 1967 Note on the Kerr Metric and Rotating Masses {\it J. Math. Phys.} {\bf 8} 1477 

\bibitem{dlc} de la Cruz V and Israel W 1968 Spinning Shell as a Source of the Kerr Metric {\it Phys. Rev.} {\bf 170} 1187 

\bibitem{pf} Pfister H and Braun K H 1985 Induction of correct centrifugal force in a rotating mass shell {\it Class. Quantum Grav.} {\bf 2} 909  

\bibitem{pf2} Pfister H and Braun K H 1986 A mass shell with flat interior cannot rotate rigidly {\it Class. Quantum Grav.} {\bf 3} 335  

\bibitem{abramo} Abramowicz M A and Prasanna A R 1990 Centrifugal-force reversal near a Schwarzschild black hole {\it Mon. Not. R. Astron. Soc.} {\bf 245} 720  

\bibitem{Friedman} Friedman J L 1978 Ergosphere instability {\it Commun. Math. Phys.} {\bf 63} 243

\bibitem{Comins} Comins N and Schutz B F 1978 On the ergoregion instability {\it Proc. R. Soc. Lond.} A {\bf 364} 211 

\bibitem{Schutz} Schutz B F and Comins N 1978 On the existence of ergoregions in rotating stars {\it Mon. Not. R. Astron. Soc.} {\bf 182} 69 

\bibitem{shin} Yoshida S and Eriguchi E 1996 Ergoregion instability revisited -- a new and general method for numerical analysis of stability {\it Mon. Not. R. Astron. Soc.} {\bf 282} 580

\bibitem{Bardeen} Bardeen J M, Press W H and Teukolsky S A 1972 Rotating black holes: Locally nonrotating frames, energy extraction, and scalar synchrotron radiation { \it Astrophys. J.} {\bf 178} 347 
  
  
\end{thebibliography}
\end{document}